\documentclass{aa}  

\usepackage{txfonts}
\usepackage{hyperref}
\usepackage[flushleft]{threeparttable}
\usepackage{graphicx} % Required for including images
\usepackage{subcaption} % Required for creating subfigures

\usepackage[dvipsnames]{xcolor} % REMOVE LATER
\usepackage{color}
\definecolor{mhi}{rgb}{0.6,0.0,0.6}

\usepackage{subcaption}
\usepackage{multirow}
\begin{document}

\title{GAMA~526784: the progenitor of a globular cluster-rich ultra-diffuse galaxy?}

\subtitle{I. Star clusters, stellar body and ionised gas properties}

% \subtitle{I. Overviewing the $\kappa$-mechanism}

\author{Maria Luisa Buzzo\inst{1,2,*}
\and Michael Hilker\inst{1}
\and Anita Zanella\inst{3}
\and Katja Fahrion\inst{4}
\and Richard M. McDermid\inst{5,6}
\and Remco van~der~Burg\inst{1}
\and Marco Mirabile\inst{7,8,1}
}

\institute{European Southern Observatory, Karl-Schwarzschild-Strasse 2, 85748 Garching bei M\"unchen, Germany 
\and Centre for Astrophysics and Supercomputing, Swinburne University, John Street, Hawthorn VIC 3122, Australia 
\and INAF - Osservatorio di Astrofisica e Scienza dello Spazio di Bologna, Via Gobetti 93/3, I-40129, Bologna, Italy
\and Department of Astrophysics, University of Vienna, T\"urkenschanzstraße 17, 1180 Wien, Austria
\and Australian Astronomical Optics (AAO), Macquarie University, NSW 2109, Australia
\and Macquarie University Astrophysics and Space Technloges, NSW 2109, Australia
\and INAF Osservatorio Astronomico d’Abruzzo, Via Maggini, 64100 Teramo, Italy
\and Gran SassoScience Institute, Via Francesco Crispi 7, L’Aquila, 47100, Italy
\\ *\email{luisa.buzzo@gmail.com}
         }

   \date{Received XX; accepted YY}

% \abstract{}{}{}{}{} 
% 5 {} token are mandatory
 
  \abstract
  % context heading (optional)
{Ultra-diffuse galaxies (UDGs) are an intriguing population of galaxies. Despite their dwarf-like stellar masses and low surface brightness, they have large half-light radii and exhibit a diverse range of globular cluster (GC) populations. Some UDGs host many GCs while others have none, raising questions about the conditions under which star clusters form in dwarf galaxies. GAMA~526784, an isolated UDG with both an old stellar body and an extended star-forming front, including many young star clusters, provides an exceptional case to explore the link between UDG evolution and star cluster formation.}
{This study investigates the stellar populations, star clusters, ionised gas properties, and kinematics of GAMA~526784, focusing on the galaxy's potential to form massive GCs and its connection to broader UDG formation scenarios.}
{Imaging from HST and Subaru/HSC, alongside MUSE spectroscopy, were used to analyse the galaxy's morphology, chemical composition, and kinematics. A combination of SED fitting and full spectral fitting was applied.}
{GAMA~526784's central stellar body exhibits a low-metallicity ([M/H] $\sim -1.0$ dex) and an old age ($t_M \sim 9.9$ Gyr), indicative of a quiescent core. The outskirts are much younger ($t_M \sim 0.9$ Gyr), but slightly more metal-poor ([M/H] $\sim -1.2$ dex). The stellar kinematics show low velocity dispersions ($\sim 10$ km s$^{-1}$) and a coherent rotational field, while the ionised gas exhibits higher dispersions (reaching $\sim 50$ km s$^{-1}$), a misaligned rotation axis ($\sim 20^\circ$) and localised star formation, what could be suggestive of a recent interaction. The young star clusters span ages of $8-11$ Myr and masses of $\log(M_{\star}/M_{\odot}) \sim 5.0$, while the old GC candidates have $\sim 9$ Gyr and stellar masses of $\log(M_{\star}/M_{\odot}) \sim 5.5$.}
{GAMA~526784's properties point to interactions that triggered localised star formation, leading to the formation of young star clusters. Future observations of its molecular and neutral gas content will help assess its environment, the trigger of this star-forming episode, and explore its potential to sustain star formation.}


   \maketitle
%
%-------------------------------------------------------------------

\section{Introduction}

Star cluster formation in dwarf galaxies has long been a subject of intense study, with open questions about why some dwarf galaxies form and host globular clusters (GCs) while others do not \citep[see e.g.][]{Billett_02,Weidner_04,Larsen_14, Kruijssen_14, Adamo_20, Kruijssen_25}. This debate has gained renewed importance with the discovery of ultra-diffuse galaxies (UDGs), a population of faint, extended galaxies with dwarf-like stellar masses but sizes comparable to much larger systems. UDGs exhibit an astonishing range in their GC populations, from none to dozens, making them key testbeds for understanding the conditions under which star clusters form in dwarf galaxies and their relationship to galaxy evolution \citep{vanDokkum_15, Beasley_16, Peng_06}.

UDGs challenge conventional galaxy formation theories. Despite their low surface brightnesses ($\mu_{g,0} > 24$ mag arcsec$^{-2}$), they are spatially extended, with half-light radii ($R_{\rm e} > 1.5$ kpc) similar to those of more massive galaxies \citep{vanDokkum_15, Koda_15}. The physical mechanisms behind their large sizes remain debated, and it is unclear whether these same processes also explain their diverse GC populations. 

\begin{figure*}
    \centering
    \includegraphics[width=\textwidth]{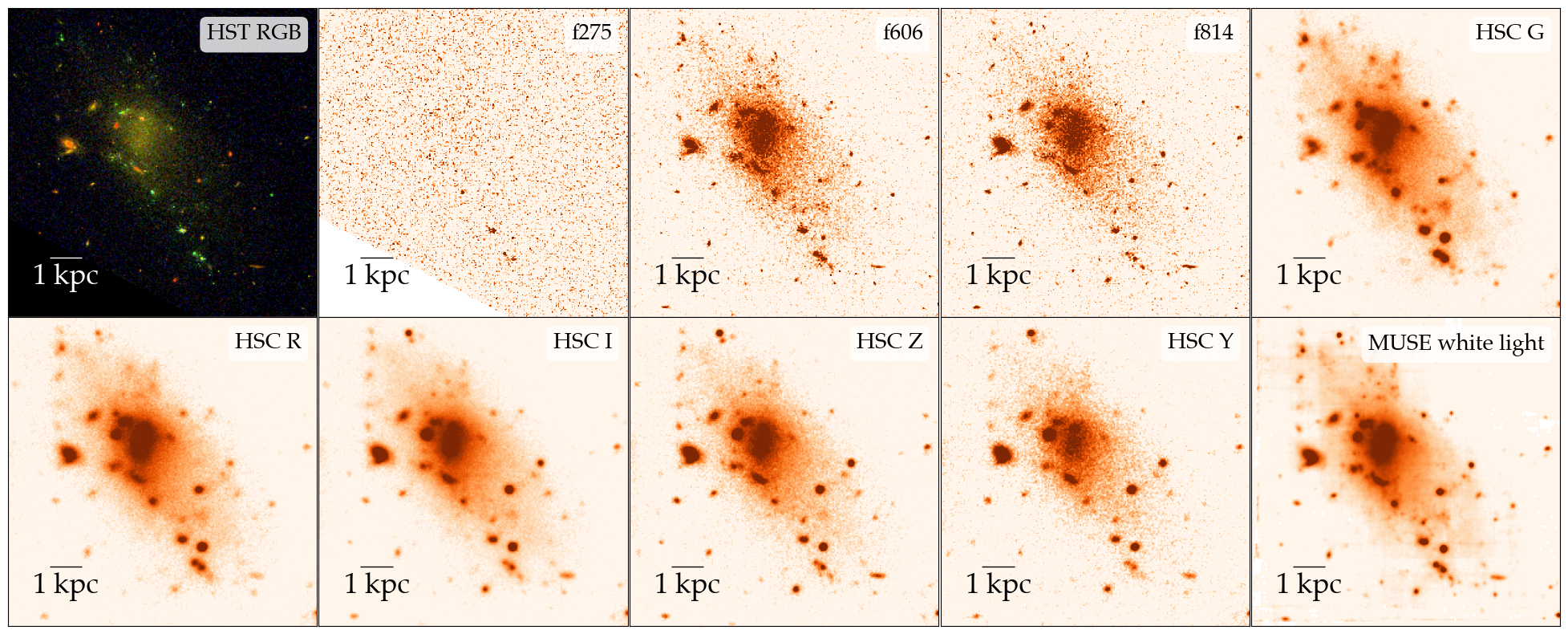}
    \caption{Postage stamps of GAMA~526784 in all wavelengths used in this study. \textit{Top left to bottom right:} HST RGB image created from the three available HST bands; HST WFC3/UVIS F275W; HST ACS F606W; HST ACS F814W; HSC $g$-band; HSC $r$-band; HSC $i$-band; HSC $z$-band; HSC $Y$-band; VLT/MUSE white-light image. In all panels, the sky background has been subtracted, and the scaling is uniform to allow reliable comparisons. Each panel is 1 sq. arcmin. North is up, and east is to the left.}
    \label{fig:data_all}
\end{figure*}

Several formation pathways have been proposed to explain UDGs \citep{Forbes_20, Buzzo_22b, Ferre-Mateu_23, Buzzo_24, Buzzo_25a}. Broadly, they seem to split into two main types: (1) those that resemble regular dwarf galaxies, and (2) those with anomalously rich GC systems \citep[see e.g.,][]{Forbes_Gannon_24, Forbes_25}. The first type, often referred to as ``puffed-up dwarfs'', is thought to originate from classical dwarfs that became more extended due to internal or external processes, such as tidal heating, feedback-driven expansion, or formation within high-spin halos \citep{Amorisco_16, Rong_17, diCintio_17, Jiang_19, Benavides_23, Sales_20}. The second type, sometimes called ``failed galaxies'', is often characterised by an unusually high number of GCs or a high GC mass fraction relative to the galaxy's stellar mass. These systems are hypothesised to have experienced an early burst of intense, high-density star formation, leading to the formation of many GCs before further star formation was suppressed, leaving behind a diffuse stellar component alongside a vast GC population \citep{Adamo_20, Cook_23, Bastian_20}. 

While the formation of GC-rich UDGs in dense environments has some explanations, such as tidal stripping or heating \citep{Carleton_21, Pfeffer_24}, their existence in complete isolation suggests that other formation mechanisms may also be at play. These could include high-speed galaxy collisions that led to compact, massive star-forming regions \citep[e.g.,][]{Silk_19, vanDokkum_22}; interactions between gas-rich dwarfs and more massive systems, triggering strong bursts of star formation that either completely destroy the dwarfs or exhaust their gas supplies \citep[e.g.,][]{Fensch_19b, Elmegreen_18}; among others.

The most direct way to study the link between UDGs and star cluster formation is to observe UDGs currently forming star clusters. However, this is challenging, as most known star-forming UDGs today have low star formation rates and do not host significant GC populations \citep{Jones_23, Leisman_17}. This makes the discovery of actively star-forming UDGs with young star clusters particularly valuable for understanding the early stages of GC formation.

In this study, we focus on GAMA~526784, an isolated UDG in one of the GAMA fields \citep{Driver_11}. First identified by \citet{Prole_19}, this galaxy provides a rare opportunity to observe star cluster formation in a UDG environment. It lies at a distance of $\sim39$ Mpc (assuming its radial velocity follows the Hubble flow) and it may be classified as an ultra-diffuse galaxy if we consider the uncertainties, lying at the very upper bound of the surface brightness criterion, with an effective radius of $2.5 \pm 0.1$ kpc and a central surface brightness of $23.7 \pm 0.3$ mag arcsec$^{-2}$. Its total stellar mass is $\log(M_{\star}/M_{\odot}) = 8.34 \pm 0.30$. Notably, GAMA~526784 hosts both an old, extended stellar body and a collection of luminous, star-forming clumps arranged in a string-like configuration spanning over 5 kpc (see Figure~\ref{fig:data_all}). 

To explore the possible formation scenarios of the galaxy and its star clusters, we present deep imaging and spectroscopy of GAMA~526784, combining space- and ground-based observations to study the kinematics, metallicities, and ages of both its stellar body and star-forming clumps. This work is the first in a series on GAMA~526784. A follow-up study will include ALMA and GBT observations to analyse the molecular and neutral gas content of the galaxy, as well as its large-scale environment. By combining multiwavelength observations, this system provides a benchmark for testing whether isolated, star cluster-rich UDGs could evolve into GC-rich UDGs and whether they follow distinct evolutionary pathways. 

\section{Data}

In this work, we combine multiwavelength optical imaging and integral field spectroscopy (IFS) to explore the galaxy's stellar populations, star-forming regions, and ionised gas properties. Specifically, we employ imaging data from the Hubble Space Telescope (HST, spatial resolution of 0.05 arcsec) and the Hyper Suprime-Cam (HSC, pixel scale of 0.17 arcsec/pixel and median seeing of 0.6 arcsec in the $i$ band) on the Subaru Telescope, alongside IFU spectroscopy from the Multi Unit Spectroscopic Explorer (MUSE, pixel scale of 0.2 arcsec/pixel and median seeing of 0.59 arcsec for our observations) on the Very Large Telescope (VLT). Figure~\ref{fig:data_all} showcases the imaging data used in this study.

\subsection{Hubble Space Telescope Observations}

\begin{figure*}
    \centering
    \includegraphics[width=\textwidth]{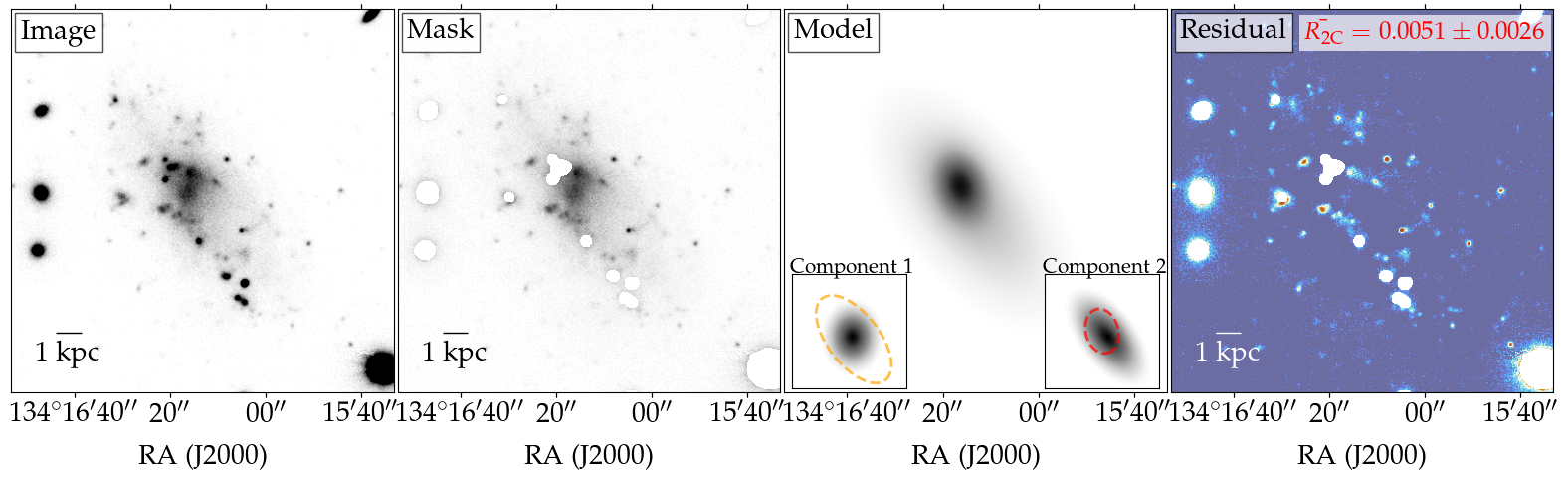}
    \caption{\texttt{GALFITM} two-component model of GAMA~526784. \textit{Columns from left to right:} HSC $g$-band image, segmentation mask from \texttt{SExtractor}, fitted galaxy model with subcomponents highlighted in inset panels, and residual (image$-$model). The subcomponents are both centred at the coordinates of the inner component to facilitate direct comparison. For reference, the one effective radius ellipses of the complementary component are overlaid in each panel. The median residuals after removing the contribution from masked sources are $0.0051 \pm 0.0026$. The inner and outer components are distinctly visible, with the smaller, more circular inner component offset and twisted relative to the elongated outer component.}
    \label{fig:galfitm}
\end{figure*}

Observations of GAMA~526784 were carried out using the Wide Field Camera 3 (WFC3/UVIS) in the F275W band with a total exposure time of 2336\,s over one orbit, and the Advanced Camera for Surveys (ACS/WFC) in the F606W and F814W bands with total exposure times of 2080\,s and 2064\,s, respectively, across two orbits (PI: van der Burg, Program ID: 16186). Each orbit was divided into four separate observations to facilitate dithering and mitigate cosmic ray contamination. The data were processed and combined using the \texttt{AstroDrizzle} task from \texttt{DRIZZLEPAC}, employing inverse-variance weight maps for optimal signal-to-noise (S/N) weighting. 

\subsection{Hyper Suprime-Cam Observations}

GAMA~526784 was also imaged as part of the Hyper Suprime-Cam Subaru Strategic Program (HSC-SSP; \citealt{Aihara_18}). Observations were conducted in five broadband filters: $g$, $r$, $i$, $z$, and $Y$. The HSC data provide wide-field imaging that complements the high-resolution HST data, enabling a detailed study of the galaxy and its environment (see Figure~\ref{fig:data_all}).

\subsubsection{Photometric Analysis with GALFITM}

To extract structural parameters, we employed \texttt{GALFITM} \citep{Haussler_13,Vika_13} for simultaneous multi-band fitting of the HSC imaging data. This approach leverages information from both high and low S/N bands, producing consistent and well-constrained photometric parameters across the entire dataset.

Due to the irregular shape of the galaxy, we tested both a single-S\'ersic and a two-component model to identify its most appropriate representation. 
In both the single- and two-component model, our \texttt{GALFITM} setup allows magnitudes to vary freely across bands, while parameters such as the effective radius ($R_{\rm e}$), S\'ersic index ($n$), axis ratio ($b/a$), and position angle (PA) are held constant across all bands for each component, but can vary on a component to component basis.
Synthetic PSFs for each band were generated using \texttt{PSFex} \citep{Bertin_02}. Interlopers, including foreground and background objects, were masked using segmentation maps produced with \texttt{SExtractor} \citep{Bertin_02}.

Figure~\ref{fig:galfitm} shows the result of our two-component fit to the HSC $g$-band image. The two components are clearly visible in both the original and model images. For comparison, the single-Sérsic fit using \texttt{GALFITM} is presented in Appendix~\ref{sec:appendix_galfitm}. To assess the robustness of the fit, we follow the approach of \cite{Buzzo_21b} and compute the median residuals, finding that the two-component model yields residuals closer to zero ($\bar{R}_{\mathrm{2C}} = 0.0051 \pm 0.0026$) than the single-component model ($\bar{R}_{\mathrm{1C}} = 0.0095 \pm 0.0021$). While we acknowledge that the two-component model has more degrees of freedom, the improvement is not only quantitative but also visually evident in the residual maps (Figures~\ref{fig:galfitm} and \ref{fig:galfitm_1comp}), where the central light profile is clearly underfit by the single Sérsic model. While the statistical significance of this improvement may be limited, we argue that the two-component model provides a more physically meaningful representation of the galaxy’s structure. This interpretation is further supported by the analysis of their stellar populations in Section~\ref{sec:SED_fitting}.
Another caveat to the two-component model is that it assumes axisymmetric subcomponents and does not allow for alternative possibilities, such as a triaxial morphology that can also produce isophote twists similar to those observed in the inset panels in Figure~\ref{fig:galfitm}. Thus, while the two-component model provides a physically plausible description of the galaxy’s structure, it should be considered one of several possible interpretations.

\subsection{VLT/MUSE Spectroscopy}

VLT/MUSE data on GAMA~526784 were obtained during dark time on 30 December 2021 (PI: Prole, ESO programme ID 108.21ZY.001). A total of three hours of on-target exposure were divided into three observing blocks (OBs), each comprising four dithered sub-exposures. The observations were conducted with adaptive optics in Wide Field Mode, yielding a $1 \times 1$ arcmin$^2$ field of view with a spatial sampling of $0.2 \times 0.2$ arcsec$^2$. The wavelength coverage spans 4800 to 9300 \AA, with a spectral resolution ranging from 69 km s$^{-1}$ (at 5000 \AA) to 46 km s$^{-1}$ (at 7000 \AA) \citep{Bacon_17,Guerou_17,Emsellem_19}. 

The data were reduced using the MUSE pipeline (v2.8.5) \citep{Weilbacher_16} within the \texttt{ESOREFLEX} environment \citep{Freudling_13}. The standard reduction steps included bias and flat-field corrections, wavelength calibration, and illumination correction. 

Sky subtraction was optimised following the method described in \cite{Iodice_23} and \cite{Hartke_25}. After an initial reduction, a custom sky mask was generated using \texttt{SExtractor}, which excluded spaxels associated with the galaxy, clusters, and foreground/background objects. This mask was then applied in subsequent reduction steps, with a sky fraction of 75\% for estimating the background. Finally, residuals were cleaned with \texttt{ZAP} \citep{Soto_16} to produce the final data cube. The MUSE white-light image is shown in the last panel of Figure~\ref{fig:data_all}.

\begin{figure*}
    \centering
    \begin{subfigure}[t]{0.5\textwidth}
        \centering
        % \textbf{Inner Component} % The title above the subfigure
        \par\medskip % Space between the title and the subfigure
        \includegraphics[width=1.3\textwidth]{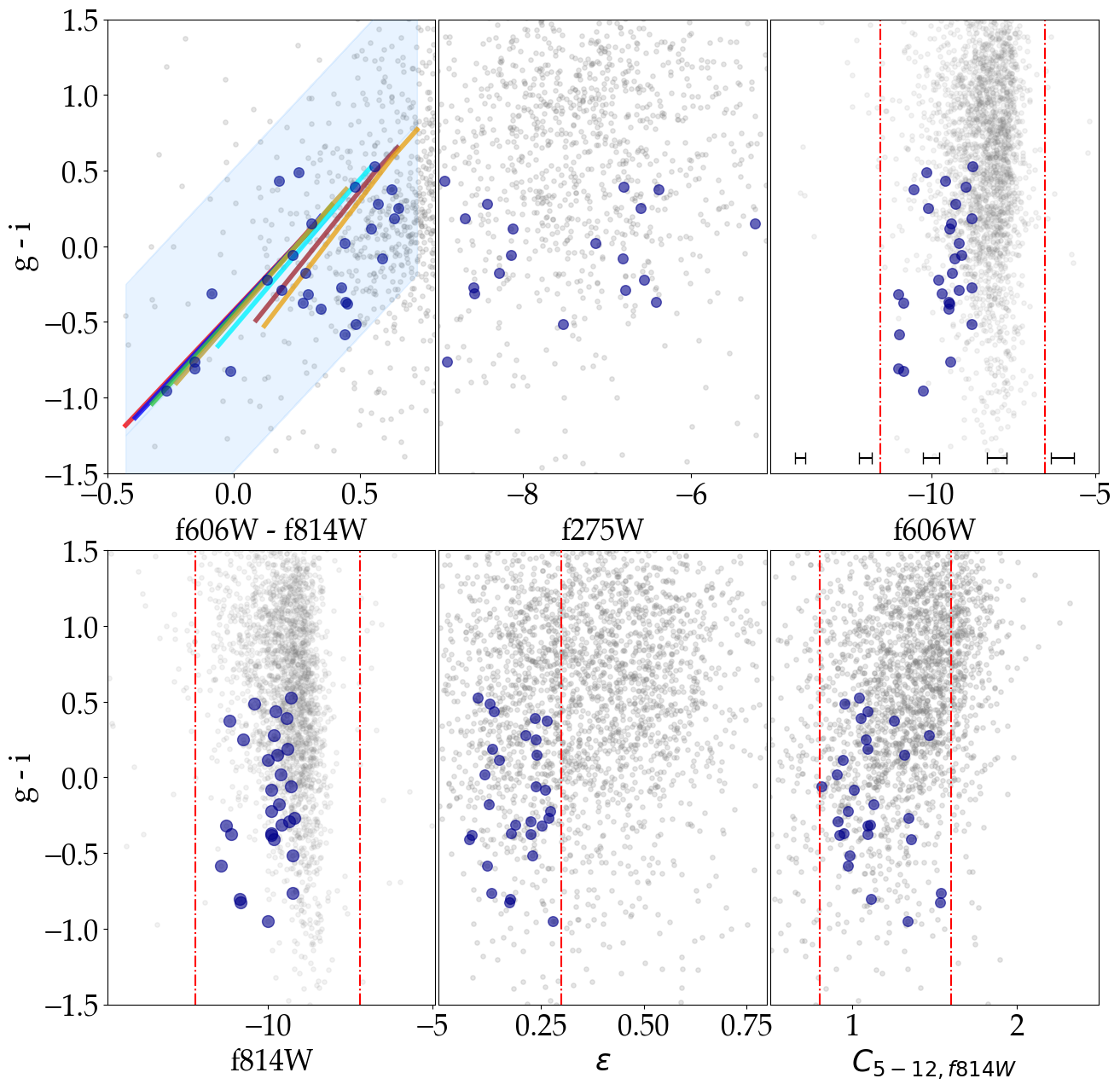}
    \end{subfigure}
    \hfill
    \begin{subfigure}[t]{0.35\textwidth}
        \centering
        % \textbf{Outer Component} % The title above the subfigure
        \par\medskip % Space between the title and the subfigure
        \vspace{0.8cm}
        \includegraphics[width=1.5\textwidth,angle=90]{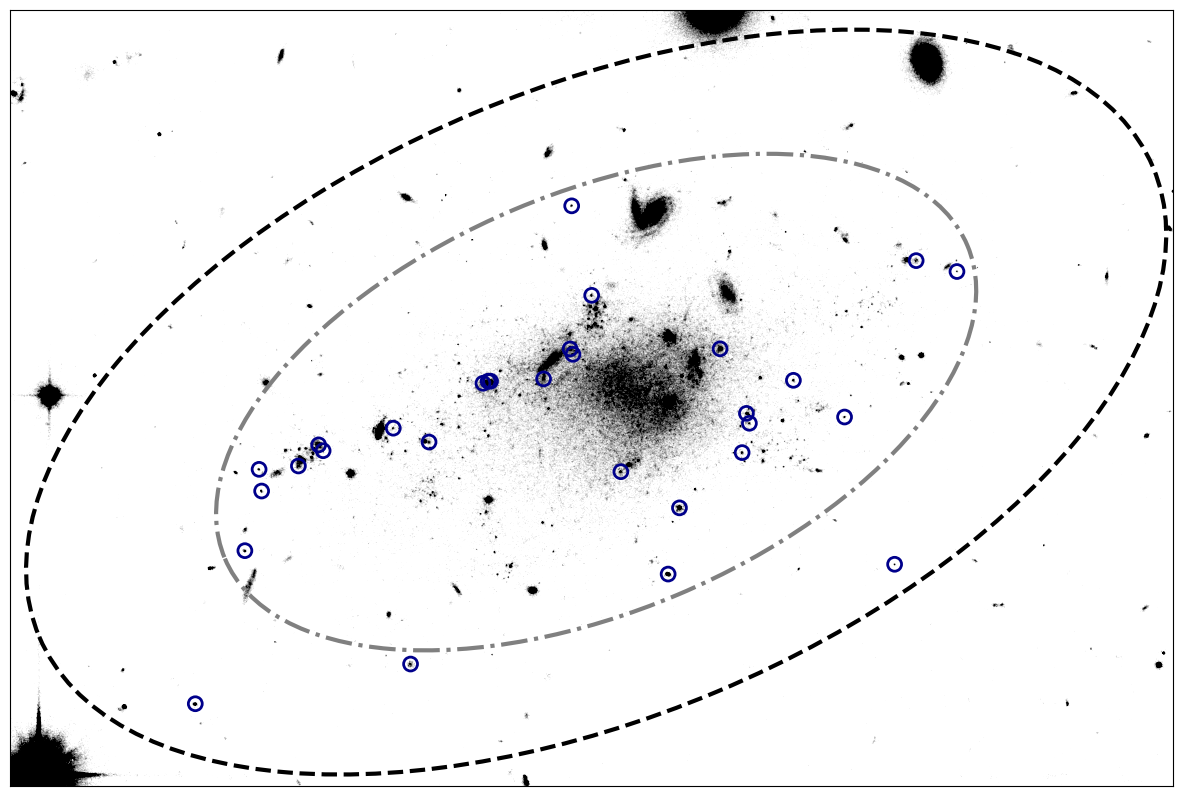}
    \end{subfigure}
    \caption{Star cluster selection criteria based on expected colours, magnitudes, ellipticity and concentration. In all panels, grey dots represent the detections from \texttt{SExtractor}. \textit{Top left:} colours predicted from SSP models are shown as continuous lines, with different colours corresponding to metallicities ranging from $-2.0$ to 0.2 dex. The extent of the lines reflects the age range (0.1 to 14 Gyr). We fit a linear relation to these predictions and define an error margin of 0.5 ($\sim 3\sigma$) mag to account for uncertainties in both the models and the observed colours. The blue-shaded region shows this allowed space. \textit{Top middle, top right and bottom left:} colour–magnitude distribution of detected sources. We applied magnitude cuts in the $F606W$ and $F814W$ bands based on the expected peak of the luminosity function for young star clusters, allowing for sources 3$\sigma$ brighter and 1$\sigma$ fainter than the peak. The error bars at the bottom of the $F606W$ panel show the typical uncertainties of sources at the respective magnitudes. \textit{Bottom middle:} Ellipticity cuts to select unresolved or nearly circular sources. \textit{Bottom right:} Concentration parameter defined as the magnitude difference in 5 and 12 aperture radii. \textit{Rightmost panel:} Star cluster candidate distribution on top of the combined F606W+F814W image, the dash-dotted line marks $2~R_e$, while the dashed line marks $3~R_e$. After applying these selection criteria, we identify 29 star cluster candidates within $3~R_{\rm e}$ of GAMA~526784.}
    \label{fig:GC_selection}
\end{figure*}

\section{Star Cluster Detection and Selection}

\label{sec:cluster_selection}

To detect and measure the photometric properties of star clusters, we used \texttt{SExtractor} \citep{Bertin_02} in dual-image mode. The detection image was created by combining the $F606W$ and $F814W$ bands, while the photometry was measured on the individual bands. The HSC bands were reprojected onto the same pixel scale as the HST imaging using \texttt{Montage} \citep{Jacob_10, Berriman_17}, enabling consistent photometry across all bands. Aperture magnitudes were extracted using a 5-pixel diameter aperture, suitable for identifying unresolved or marginally resolved star clusters. This aperture ensures a clean point source selection with fewer background galaxy contaminants, but comes at the cost of possibly missing a fraction of star clusters with larger radii that are partially resolved. Background subtraction was performed locally, using a small background cell size (\texttt{BACK\_SIZE} = 24 pixels), and photometric errors were calculated from \texttt{SExtractor}'s nominal error estimates. Galactic extinction corrections (E(B-V) = 0.034) were applied using the reddening law of \citet{Fitzpatrick_99} and the recalibration by \citet{Schlafly_11}.

\subsection{Selection Criteria}

The selection of star clusters was based on a combination of morphological, photometric, and colour-based criteria.  
We employed simple stellar population (SSP) models consistent with the expected stellar populations of star clusters to guide our selection. These SSP models were constructed using the Flexible Stellar Population Synthesis package \citep[FSPS;][version 0.4.2]{Conroy_09,Conroy_10a,Conroy_10b}, utilising the Padova isochrones \citep{Marigo_07,Marigo_08}. The models allowed for metallicities in the range [M/H] = $-2.2$ to $0.0$ dex and ages $t_{\rm age}$ = 0.1--14 Gyr. The predicted colours of these SSP models are shown in the leftmost panel of Figure~\ref{fig:GC_selection}, where each coloured line corresponds to a specific metallicity, and the extent of the lines represents the age range of the SSPs.

To refine our selection, we derived a linear relationship between the $F606W - F814W$ and $g - i$ colours by fitting the SSP model predictions. A similar GC selection criterion was applied in \cite{Buzzo_23}, and the fit is illustrated in the first panel of Figure~\ref{fig:GC_selection}. The derived relation is as follows:

\begin{equation}
    (g-i) = 1.79 \times (F606W - F814W) - 0.01
\end{equation}

\begin{figure}
    \centering
    \includegraphics[width=\columnwidth]{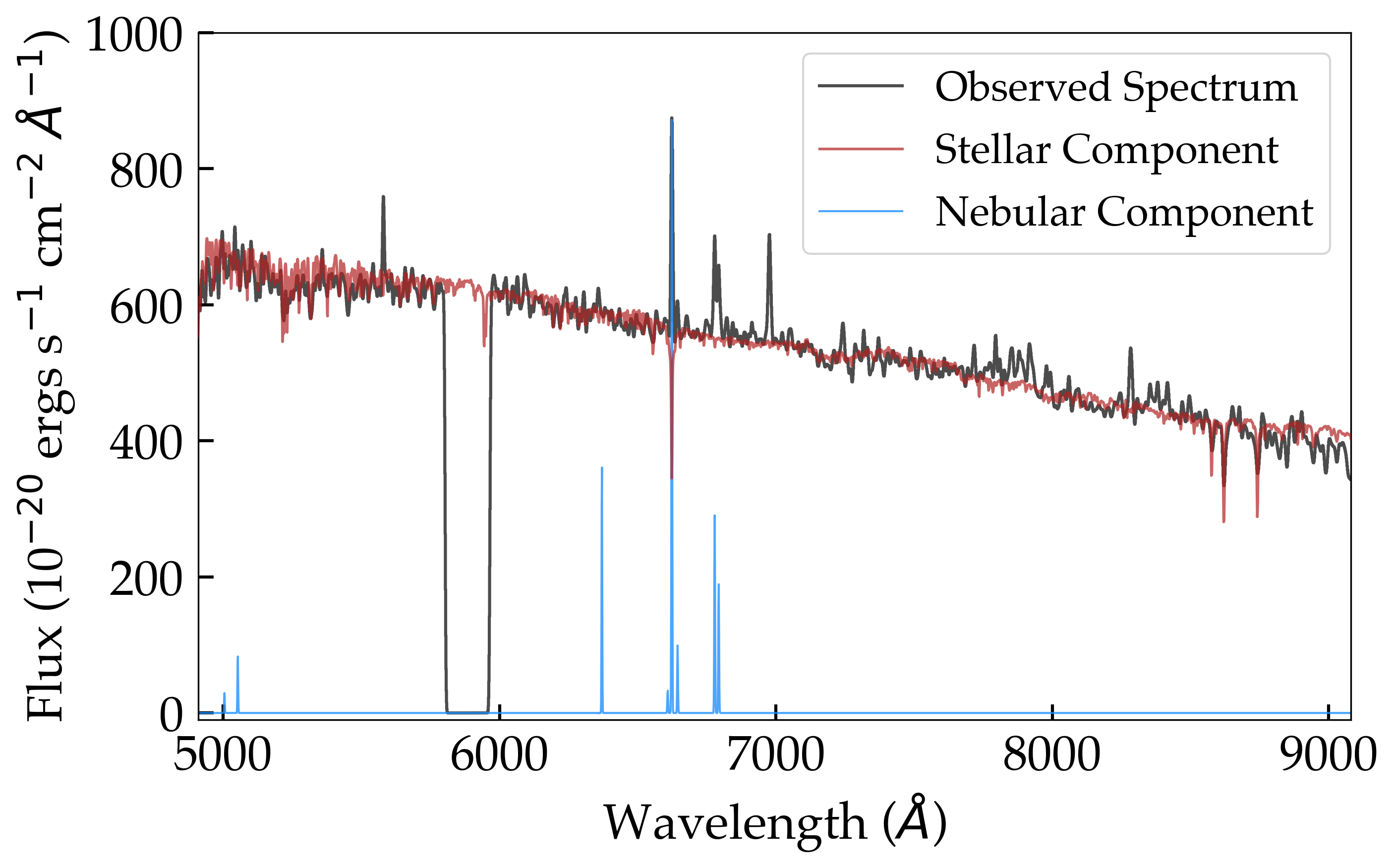}
    \caption{Observed and fitted spectrum of the central bin of GAMA~526784 using \texttt{DAP}. The panel shows the observed spectrum (grey), and the decomposed stellar (red) and nebular (blue) components. Prominent emission lines such as H$\alpha$, [NII], [SII], and [OIII], as well as stellar absorption features (e.g., H$\alpha$, Ca triplet, Mg, Fe lines), are clearly identified.}
    \label{fig:spectra}
\end{figure}

Star clusters were selected if their colours fell within $\pm0.5$ mag (approximately $3\sigma$) of the linear relation. Using the $F606W - F814W$ and $g - i$ colours ensured that we did not exclude clusters undetected in the $F275W$ band while still covering the broad colour range expected for both young and old clusters.

Using the same SSP models, we calculated a magnitude difference of 1.7 mag in $M_{F814W}$ between old GCs (aged 8--14 Gyr) and young star clusters ($\sim 10$ Myr) of the same mass. This would shift the peak of a Gaussian-shaped GCLF of $M_{F814W} = -8.2 \pm 1.0$ for an old GC system to $M_{F814W} \sim -9.9$ for young clusters (see inset in Figure~\ref{fig:starcluster_fits}). We acknowledge that the mass distribution of young clusters does not follow a Gaussian distribution but rather a power law, but the bright end might be representative of the distribution of massive young star clusters. Therefore, sources were selected if their magnitudes fell within a range of $3\sigma$ brighter and $1\sigma$ fainter than the peak of the adjusted luminosity function, assuming that fainter star clusters would dissolve over several hundred Myr. Additionally, sources with low ellipticity ($\epsilon < 0.3$) were selected to ensure they were nearly circular and consistent with being unresolved or marginally resolved at the distance of GAMA~526784.

We further refined the selection by introducing a concentration parameter \citep{Harris_96, Peng_06}, shown in Figure~\ref{fig:GC_selection}, defined as the difference in magnitude between apertures of 5 and 12 pixel radii in the F814W image. The concentration has been widely adopted as a proxy for distinguishing compact sources, such as star clusters, from extended contaminants \citep[e.g.][]{Janssens_22, Janssens_24, Tang_25, Marleau_24}. We found that genuine star cluster candidates populate a well-defined range in concentration (between $0.8 < C_{5-12,f814W} < 1.6$), consistent with expectations for unresolved or marginally resolved sources at the distance of GAMA~526784. 

Star cluster candidates were considered to belong to GAMA~526784 if they fell within $3~R_{\rm e}$ of the galaxy. For this, we assumed the $R_e$ recovered for the more elongated component, i.e., 2.9 kpc and used an elliptical aperture, assuming also the b/a of the outer component, b/a = 0.5. This aperture is larger than the traditional $2~R_{\rm e}$ commonly used in similar studies (e.g., \citealt{Beasley_16}), to account for the extended and highly elongated spatial distribution of star clusters in GAMA~526784. We note, nonetheless, that assuming three instead of two effective radii adds only three cluster candidates to our sample (out of which one is confirmed to be associated with the galaxy in Table \ref{tab:stellarpops}).

After applying these criteria, we identified 29 star cluster candidates. The properties of these clusters are shown in Figure~\ref{fig:GC_selection}, along with the cluster distribution displayed on top of the combined F606W+F814W image.
The final sample of clusters includes eight with colours consistent with old, metal-poor GCs, and the remaining are young, star-forming clusters.
In Appendix \ref{sec:number_density}, we show the star cluster number density profile, where we fit a S\'ersic profile to the star cluster distribution to better understand its profile. From this, we find a background contamination of 4.2 clusters/arcmin$^{-2}$, and a star cluster half-number radius of 1.8 $R_e$. This translates into an expected fraction of 3.8 interlopers within the three effective radii area. Out of these, 1.3 are expected to have colours similar to those of old GCs.

\section{Results}

In this section, we present the results obtained from the analysis of GAMA~526784. The results are divided into photometric and spectroscopic analyses, which encompass stellar population properties and ionised gas mapping, followed by a kinematic analysis and a stellar cluster analysis.

\subsection{Photometric Analysis: SED Fitting}
\label{sec:SED_fitting}

\begin{figure*}
    \centering
    \includegraphics[width=\textwidth]{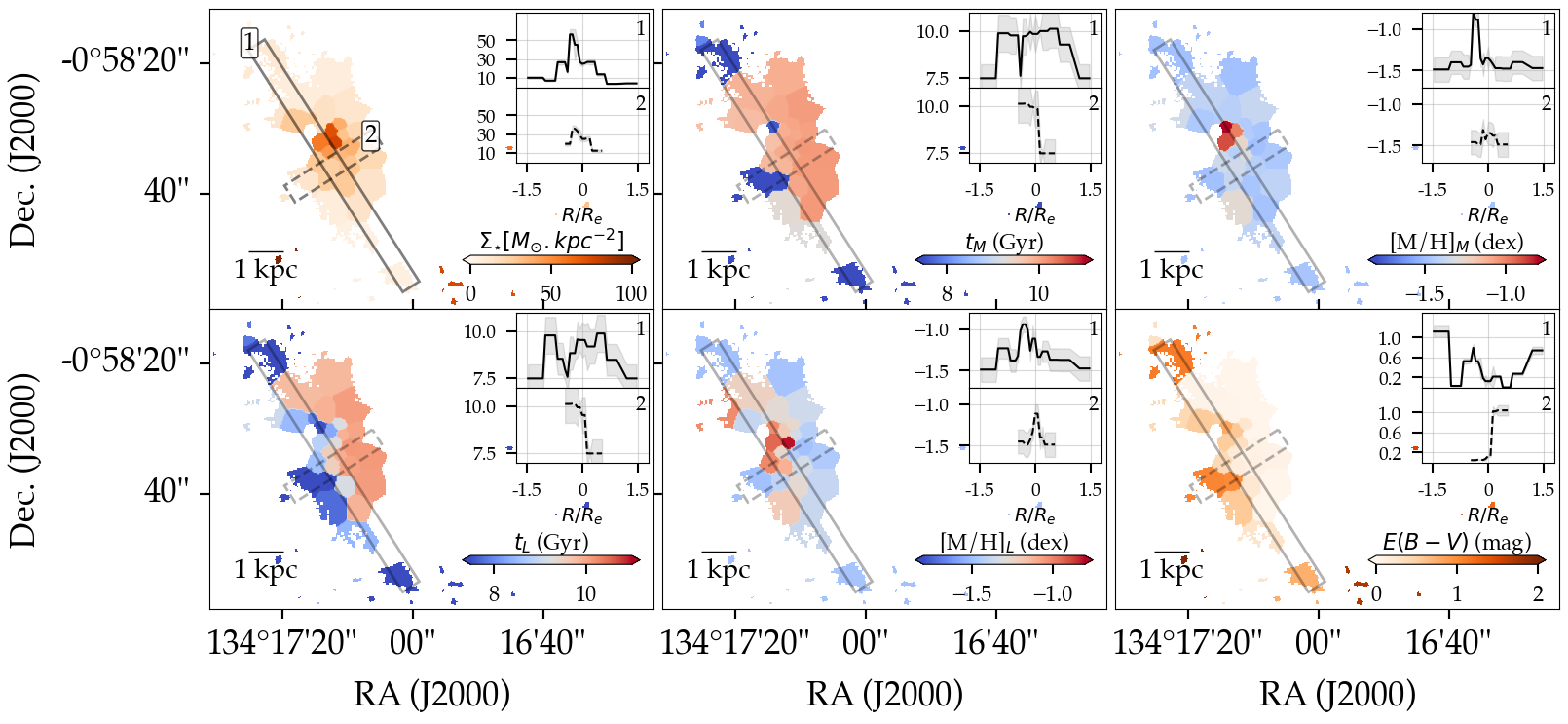}
    \caption{Spatially resolved stellar population properties of GAMA~526784 derived using the \texttt{DAP} routine. \textit{Top row, left to right:} Stellar surface mass density ($\Sigma_\star$), mass-weighted age ($t_{\mathrm{M}}$), and mass-weighted metallicity ([M/H]$_{\mathrm{M}}$). \textit{Bottom row, left to right:} Light-weighted age ($t_{\mathrm{L}}$), light-weighted metallicity ([M/H]$_{\mathrm{L}}$), and colour excess ($E(B-V)$). Inset panels show the property gradients measured along the two rectangles overlaid on the maps, with grey bands indicating the measurement uncertainties. The maps and profiles reveal a central region that is both more massive and metal-rich compared to the outskirts. Additionally, the western side of the galaxy appears younger and dustier, coinciding with the location of newly formed star clusters.}
    \label{fig:whitelight}
\end{figure*}

We performed spectral energy distribution (SED) fitting on the two structural components of GAMA~526784, recovered from the \texttt{GALFITM} decomposition, and on the single S\'ersic model of the galaxy using the Bayesian inference code \texttt{BAGPIPES} \citep{Carnall_18}. We assumed a bursty model to derive physical properties, including stellar mass, metallicity, dust attenuation, and age.  The SED fitting used the broadband HSC bands and incorporated broad priors to ensure flexibility in recovering physical properties, as shown in Figure~\ref{fig:sed_fitting}. 
The results are summarised in Table \ref{tab:data} and provide a detailed view of the stellar populations in the two components.

The inner, compact component has a stellar mass of $\log(M_\star/M_\odot) = 8.21^{+0.08}_{-0.11}$, a metallicity of $[M/H] = -1.00^{+0.23}_{-0.52}$, and an age of $9.9^{+2.4}_{-2.4}$ Gyr, indicating that this component predominantly consists of old stars. Dust attenuation is moderate in this component, with $A_V = 0.36^{+0.32}_{-0.22}$ mag. While the surface brightness of this component satisfies the UDG criteria, its effective radius ($R_{\rm e} = 0.9 \pm 0.1$ kpc) falls below the typical UDG threshold ($R_{\rm e} \geq 1.5$ kpc).  

The outer component is characterised by a stellar mass of $\log(M_\star/M_\odot) = 8.00^{+0.16}_{-0.12}$, a metallicity of $[M/H] = -1.15^{+0.30}_{-0.55}$, and an age of $0.9^{+0.9}_{-0.4}$ Gyr, indicating a predominantly young population. Dust attenuation is consistent with zero, with $A_V = 0.27^{+0.37}_{-0.21}$ mag. This component was found with \texttt{GALFITM} to have an effective radius of $2.9 \pm 0.1$ kpc and a surface brightness of $24.4 \pm 0.2$ mag arcsec$^{-2}$, thus meeting the UDG criteria.

\subsection{Spectroscopic Analysis: \texttt{DAP} Maps}

\begin{figure}
    \centering
    \includegraphics[width=0.95\columnwidth]{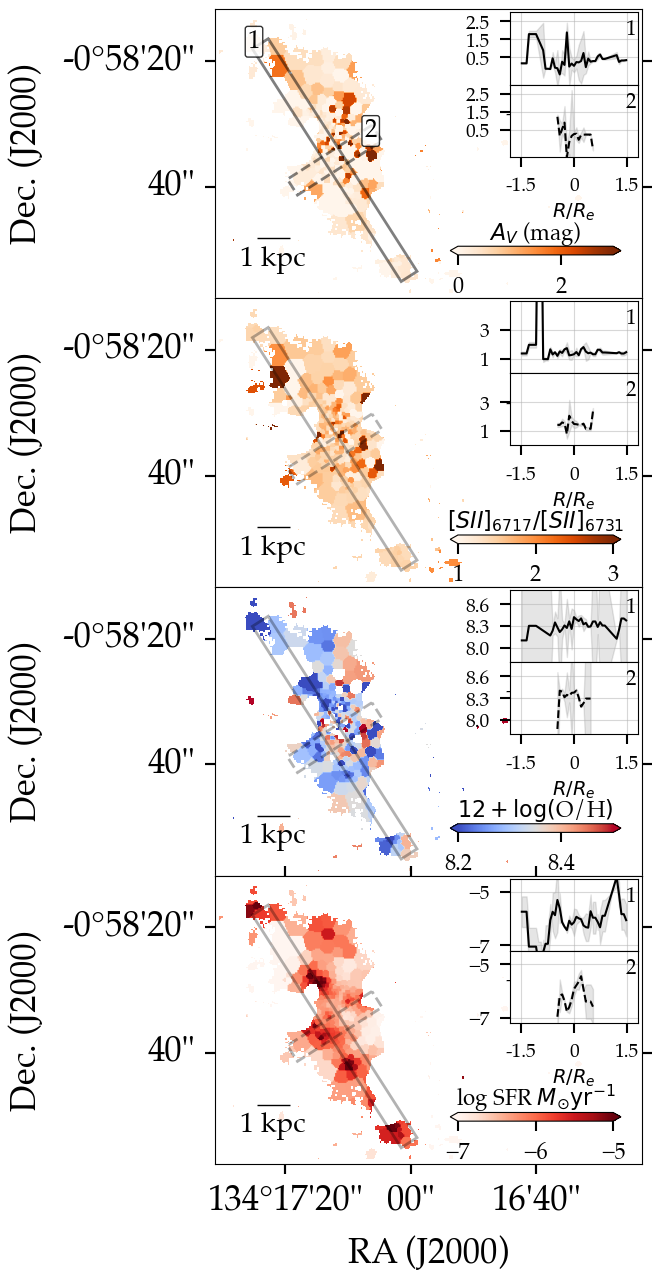}
    \caption{Ionised gas properties of GAMA~526784. \textit{First panel:} Dust attenuation ($A_V$) map derived from the Balmer decrement. \textit{Second panel:} [SII]$\lambda6716/\lambda6731$ ratio, which serves as a proxy for electron density. \textit{Third panel:} Gas-phase metallicity ($12 + \log(\mathrm{O/H})$) map estimated using the O3N2 diagnostic \citep{Marino_13}. \textit{Fourth panel:} SFR per spaxel derived from the H$\alpha$ luminosity. The inset panels show the gradients along the two rectangles in the maps, with grey bands representing measurement uncertainties. These maps reveal localised dust attenuation, an overall low [SII] ratio which may be indicative of high electron densities, low chemical enrichment in the ionised gas, and localised star formation concentrated in regions with young star clusters.}
    \label{fig:density}
\end{figure}

To analyse the MUSE spectra of GAMA~526784, we used the Data Analysis Pipeline (\texttt{DAP}; \citealt{Westfall_19,Belfiore_19}),  which is a wrapper for \texttt{pPXF} \citep{Cappellari_04,Cappellari_17} and the \texttt{Voronoi binning} code \citep[\texttt{VORBIN},][]{Cappellari_03}, and is designed to extract stellar and nebular properties from IFS data, providing a comprehensive decomposition of the light into these two key contributors. The stellar continuum was modelled using the MILES SSP templates \citep{Vazdekis_15}. For the gas component, the routine fits prominent emission lines, including H$\beta$, H$\alpha$, [NII], [SII], and [OIII], to measure the properties of the ionised gas.

To achieve robust results for both stars and gas, we applied two distinct configurations within \texttt{DAP}. For the stellar component, we defined a minimum S/N of 20 $\AA^{-1}$ in the Voronoi-binned spectra. The S/N was calculated using the continuum region between 5000 and 5600 \AA. This approach ensured high-quality stellar population fits. For the nebular emission lines, we defined a separate configuration, requiring the H$\alpha$ line to have a minimum S/N of 20 $\AA^{-1}$. This criterion effectively allowed almost all spaxels to be fit individually, as the emission lines are well-detected across the galaxy. In both cases, spaxels with S/N $\leq 0.3$\,\AA$^{-1}$ were excluded from the analysis to ensure robust measurements, and bins associated with background galaxies were masked (see red galaxies in the first panel of Figure~\ref{fig:data_all}).

Figure~\ref{fig:spectra} illustrates the results of the \texttt{DAP} fits for the central bin of the galaxy, including the decomposition into star and nebular components. 
We find that the stars and ionised gas share the same systemic velocity, suggesting that there is no significant line-of-sight separation between the red, older stellar component and the blue, star-forming component traced by the gas. This contrasts with recently discovered systems exhibiting a similar superposition of red and blue components, which have been attributed to projection effects (e.g., \citealt{Li_25}).

\subsubsection{Stellar Population Properties}

The stellar population properties of GAMA~526784 were derived by analysing the absorption lines in the MUSE spectra using the \texttt{DAP} routine. 
Figure~\ref{fig:whitelight} presents spatially resolved maps of key stellar population properties for GAMA~526784, including stellar surface mass density ($\Sigma_\star$), mass-weighted age ($t_{\mathrm{M}}$), mass-weighted metallicity ([M/H]$_{\mathrm{M}}$), light-weighted age ($t_{\mathrm{L}}$), light-weighted metallicity ([M/H]$_{\mathrm{L}}$), and colour excess ($E(B-V)$). In each map, two rectangles are overlaid: one aligned with the star clusters and the other perpendicular, enabling the study of gradients across and along the structure. These rectangles are consistently used in the analysis of both stellar (Figure~\ref{fig:whitelight}) and gas (Figure~\ref{fig:density}) components.

The calculated statistics for each map (after 3$\sigma$ clipping) reveal some key trends. The stellar surface mass density map shows a centrally concentrated distribution with a median and median absolute deviation (MAD) of $12.42 \pm 8.91\,M_{\odot} \,\rm{kpc}^{-2}$, a maximum of $36.98\,M_{\odot} \,\rm{kpc}^{-2}$, and a minimum of $1.13\,M_{\odot} \,\rm{kpc}^{-2}$, indicating a significant mass concentration towards the centre and notable variation across the galaxy. The metallicity maps ([M/H]$_{\mathrm{M}}$ and [M/H]$_{\mathrm{L}}$) show shallow gradients, with slightly higher metallicities at the centre (median of $-1.44$ dex for mass-weighted and $-1.38$ dex for light-weighted), which are consistent with the results obtained with SED fitting. The variations in metallicity across the galaxy are small, as reflected by the low MAD values (0.07 and 0.16, respectively). The maximum and minimum values for metallicity are also modest, ranging from $-1.25$ to $-1.49$ dex.

The age maps exhibit a mild radial gradient, with younger stellar populations in the western regions, which coincide with areas of active star formation (see Figure~\ref{fig:density}). The mass-weighted age map has a median of $9.80 \pm 0.45$ Gyr, while the light-weighted age map shows a more pronounced gradient, with a median of $9.05 \pm 1.53$ Gyr. This highlights the presence of younger populations in the west and older populations towards the centre. Finally, the colour excess map (median $0.27 \pm 40$ mag) indicates overall dust attenuation that is consistent with zero, though some localised regions, particularly in the west, show more significant reddening, reaching up to $E(B{-}V) = 1.26$\,mag.

While a sharply defined two-component structure is not present in any of the maps, the observed gradients in surface mass density, age, and metallicity suggest an evolved central component surrounded by a younger, more diffuse stellar population. These trends are further illustrated by the radial profiles, which show a consistent pattern of older populations in the centre and younger populations in the outskirts, aligned with the regions of star formation. Additionally, we used the light-weighted ages to reconstruct the star formation history (SFH) of the galaxy, revealing multiple star formation episodes throughout its lifetime. The SFH is discussed in Appendix \ref{sec:SFH} and shown in Fig. \ref{fig:SFH}.

\subsubsection{Ionised Gas Properties}
\label{sec:shocks_section}

\begin{figure}
    \centering
    \includegraphics[width=0.9\columnwidth]{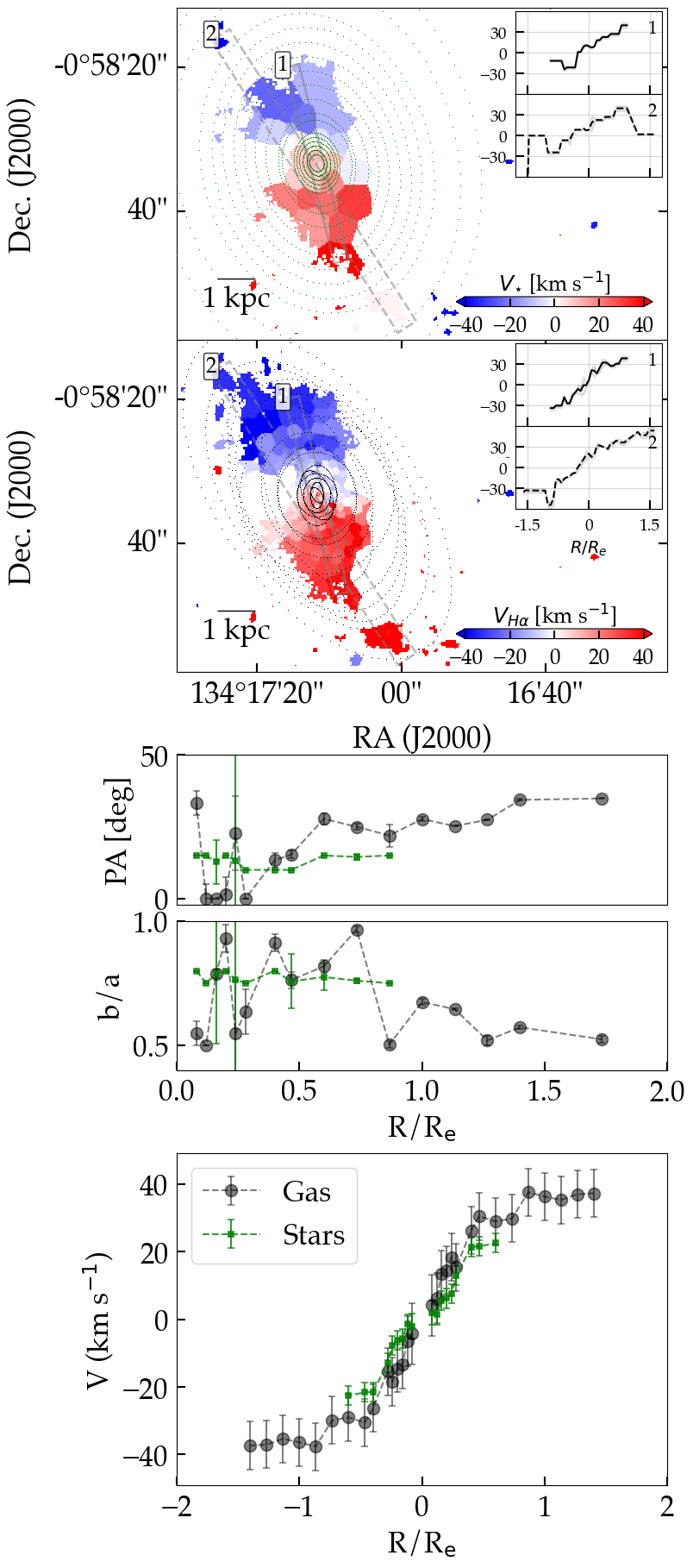}
      \caption{Velocity fields of GAMA~526784. \textit{First and second rows:} Stellar and ionised gas velocity maps, respectively. The dotted ellipses represent the best-fit ellipses from \texttt{kinemetry}, while the rectangles indicate median measurements shown in the inset panels, with the grey bands illustrating the associated uncertainties. \textit{Third and fourth panels:} Variation in the position angle and axis ratio from \texttt{kinemetry}, with the stellar component shown in green and the gas component in black. A transition is observed in the gas from the central to the outer regions, where the central component is rounder, while the outskirts become more elongated and are misaligned by approximately $20^\circ$. \textit{Fifth panel:} Best-fit velocity model from \texttt{kinemetry}. The stellar component exhibits slower rotation, with velocities around $20$\,km\,s$^{-1}$, while the gas reaches velocities of up to $40$\,km\,s$^{-1}$.}
    \label{fig:velocity}
\end{figure}

The ionised gas properties of GAMA~526784 were investigated using emission lines detected in the MUSE spectra. Figure~\ref{fig:density} presents spatially resolved maps of four key properties: (1) dust attenuation ($A_V$), derived from the Balmer decrement; (2) [SII]$\lambda6716/\lambda6731$ ratio, which serves as a proxy for electron densities; (3) gas-phase metallicity, estimated using the O3N2 diagnostic \citep{Marino_13}; and (4) SFR per spaxel derived from the H$\alpha$ luminosity map.

The dust attenuation map ($A_V$) reveals a median value of $0.29 \pm 0.64$ mag, reaching a maximum of 2.19 mag, indicating significant variation across the galaxy. Similar dust content has been reported in other UDGs with ongoing star formation \citep{Trujillo_17, Roman_Trujillo_17, Greco_18, Prole_19, KadoFong_24}.

The [SII]$\lambda6716/\lambda6731$ ratio map shows localised regions where the ratio is approximately 1.1, particularly along the string of young star clusters and in the southern part of the galaxy. This low ratio, with a median of $1.16 \pm 0.22$, suggests elevated electron densities, with values exceeding $n_\mathrm{e} \sim 400$ cm$^{-3}$ \citep{Wild_14}. The irregular spatial distribution of low [SII] ratios can be suggestive of either dense star-forming regions or shock fronts. Further observations are required to distinguish between these scenarios.

The gas-phase metallicity is fairly smooth across the galaxy, with a median of $8.31 \pm 0.11$ dex, corresponding to an overall metal-poor gas content. Using the conversion from \cite{Fraser-McKelvie_22}, the difference between gas and stellar metallicities ($\Delta Z_{g,\star}$) is 0.44 dex, suggesting that the gas is approximately 2.8 times more metal-rich than the stars. This trend is consistent with star-forming galaxies of similar stellar mass and likely reflects multiple episodes of star formation and chemical enrichment throughout the galaxy's history.

The SFR per binned spaxel was derived from the H$\alpha$ luminosity map. The median SFR is $\log(\rm{SFR}) = -6.40 \pm 0.55 \, M_{\odot} \rm{yr}^{-1}$, with values ranging from log $-7.83$ to $-4.79 \, M_{\odot} \rm{yr}^{-1}$. The integrated SFR across all spaxels is $0.01$ M$_\odot$ yr$^{-1}$ is low and consistent with the highly localised nature of star formation in GAMA~526784, concentrated in compact regions along the string of star clusters rather than being uniformly distributed across the galaxy.

\subsubsection{Velocity Maps and Kinematics}

Next, we examine the velocity fields and velocity dispersion maps of GAMA~526784, obtained using \texttt{DAP} and modelled with \texttt{kinemetry} \citep{Krajnovic_06}. \texttt{Kinemetry} approximates the velocity field using ellipses, enabling the extraction of key kinematic parameters such as the rotational velocity and velocity dispersion profiles, as well as the kinematic position angle and axis ratio, and deviations from axisymmetry.

Figure~\ref{fig:velocity} presents the velocity maps for both the stellar and ionised gas components of GAMA~526784. In this figure, we show two rectangles that depict manual measurements of the properties along their extension. These rectangles are aligned with the inner and outer components identified photometrically, rather than being perpendicular to one another as in Figures~\ref{fig:whitelight} and \ref{fig:density}, allowing for a clearer representation of the kinematic differences between the two components.

Figure~\ref{fig:velocity} shows the radial variation of the position angle (PA), axis ratio ($b/a$), and rotational velocity derived using \texttt{kinemetry} for both the ionised gas (H$\alpha$) and stellar components. The gas exhibits a clear PA misalignment of about $20^\circ$ between the inner and outer regions, while the stellar component remains relatively constant. This difference, along with the higher rotational amplitude of the gas, supports the presence of a disturbed or lopsided gas disk, possibly due to a recent interaction. For the stellar component, we show data only out to $\sim 1.0\,R_e$, where at least 70\% of each fitted ellipse contains spaxels with valid kinematic information. Beyond this radius, the fits become unreliable due to a large fraction of empty spaxels within the ellipses, which causes the velocities to default toward the systemic value. Within the reliable region, the stars rotate with an amplitude of $\sim18\,\mathrm{km\,s^{-1}}$, while the gas reaches up to $\sim40\,\mathrm{km\,s^{-1}}$ and can be traced out to $\sim 1.5\,R_e$.

\begin{figure}
    \centering
    \includegraphics[width=\columnwidth]{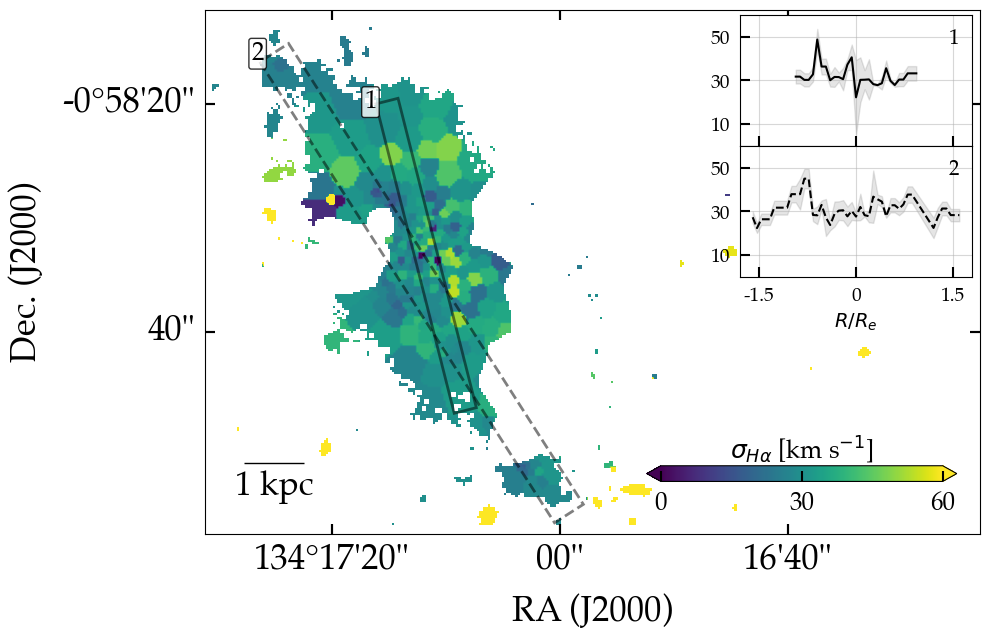}
    \caption{Gas velocity dispersion map of GAMA~526784. The map displays the velocity dispersion of the ionised gas, with rectangles indicating manual measurements shown in the inset panels. The gas velocity dispersion has a median value of $31.7 \pm 6.9$ km\,s$^{-1}$, which is significantly higher than the stellar velocity dispersion ($\sigma_{\star} = 10 \pm 6$ km\,s$^{-1}$).}
    \label{fig:dispersion}
\end{figure}

For the velocity dispersion map, only the gas component is shown in Figure~\ref{fig:dispersion}, as the stellar velocity dispersion across all Voronoi-binned spaxels is below the nominal resolution of MUSE. To derive a reliable measurement, we collapsed all spaxels and computed an integrated stellar velocity dispersion with high signal-to-noise ratio (S/N). Following the method of \cite{Iodice_23}, which demonstrates that the reliability of velocity dispersion measurements increases with S/N even below the instrumental resolution, we found a global stellar velocity dispersion of $10 \pm 6$\,km\,s$^{-1}$ at an S/N of $46$\,$\AA^{-1}$. This value aligns with expectations for a dark matter (DM)-deficient system. In contrast, a system embedded in a normal DM halo would exhibit a velocity dispersion of $\sim 25$\,km\,s$^{-1}$ \citep{Danieli_19, Shen_23, Buzzo_25b}. However, since both scenarios (being DM-deficient or residing in a normal DM halo) predict velocity dispersions that are below the nominal resolution of MUSE and differ by less than $3\sigma$, we cannot conclusively determine the DM content of the galaxy.

In contrast to the stars, the gas component shows significantly higher velocity dispersions, reaching values up to $\sim 50$\,km\,s$^{-1}$ on a spaxel-by-spaxel basis (see Figure~\ref{fig:dispersion}). The velocity dispersion of the ionised gas has a median value of $31.7 \pm 6.9$\,km\,s$^{-1}$, with a maximum of $51.4$\,km\,s$^{-1}$ and a minimum of $13.8$\,km\,s$^{-1}$. This elevated dispersion, much higher than that of the stellar component, may be attributed to shocks or the aftermath of a recent interaction.

\subsection{Analysis of Stellar Clusters in GAMA~526784}

\begin{figure*}
    \centering
    \includegraphics[width=\textwidth,trim=0cm 0.6cm 0cm 0.6cm,clip]{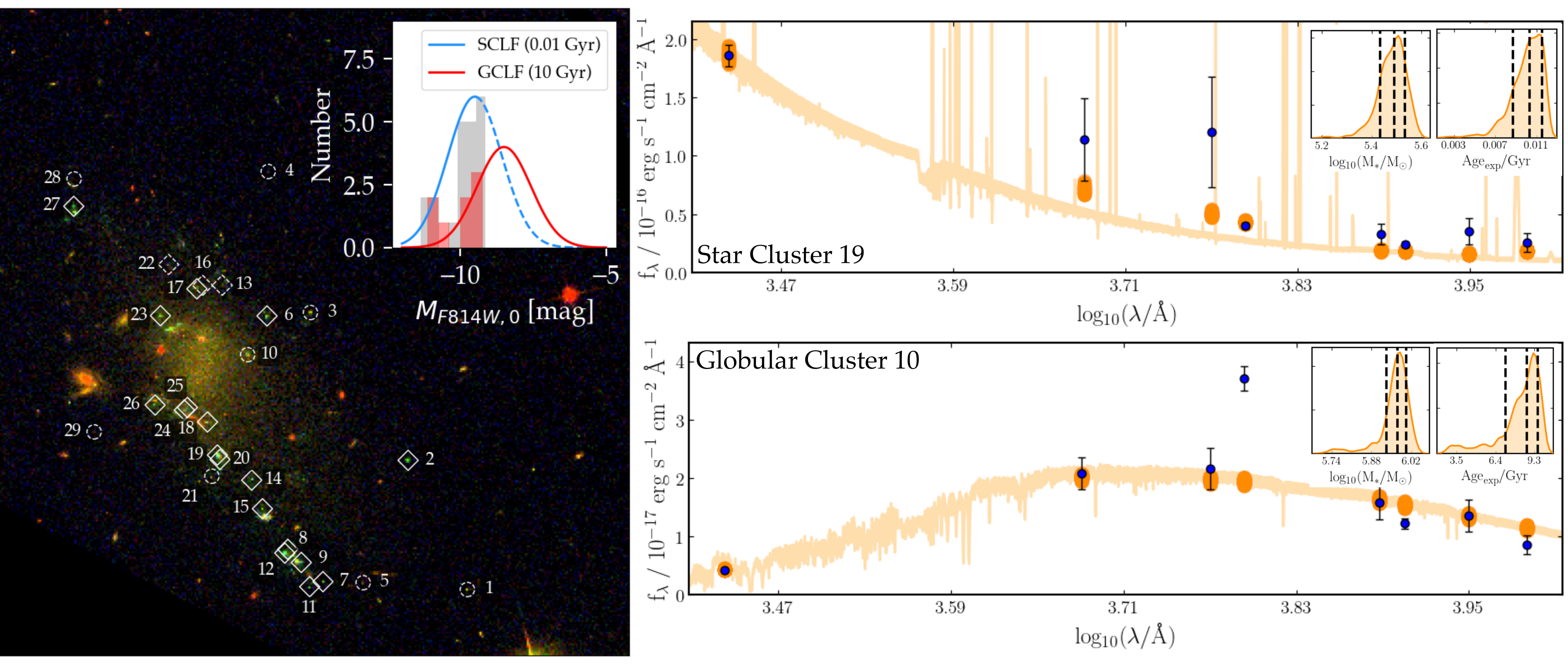}
    \caption{Stellar clusters associated with GAMA~526784. \textit{Left:} HST RGB image with the identified clusters marked and numbered in white. Old clusters are shown as circles, while young ones are diamonds. Candidates have open markers and confirmed clusters have closed ones. The inset histogram on the left compares two theoretical luminosity functions: the red curve is the typical GCLF for old globular clusters, peaking at $M_{F814W} = -8.2$ mag, while the blue curve represents the luminosity function corrected for the younger clusters' ages, with a peak at $M_{F814W} = -9.9$ mag. The grey histogram shows the magnitude distribution of the young stellar clusters, while the red one shows the old clusters. \textit{Right:} \texttt{BAGPIPES} fits for a representative star-forming cluster (number 19) and an old globular cluster (number 10). The star cluster has a young age of $t_\mathrm{M} \sim 10$ Myr, and a relatively high stellar mass of $\log(M_{\odot}/M_{\star}) \sim 5.5$. The globular cluster, alternatively, has an old age of $t_\mathrm{M} \sim 8.8$ Gyr, and a slightly higher stellar mass of $\log(M_{\odot}/M_{\star}) \sim 5.6$.}
    \label{fig:starcluster_fits}
\end{figure*}

We conducted a comprehensive spectrophotometric analysis of the stellar clusters identified around GAMA~526784 using \texttt{BAGPIPES} \citep{Carnall_18}, combining HST and HSC photometry with MUSE spectra. 
The RGB image in Figure~\ref{fig:starcluster_fits} illustrates the spatial distribution of the clusters: young star-forming clusters are associated with active star-forming regions, while older GCs are more uniformly distributed across the galaxy. The inset histogram compares two theoretical GC luminosity functions: the red curve represents the typical GCLF for old GCs (ages $8$--$14$\,Gyr), peaking at $M_{F814W} \sim -8.2$, while the blue curve is shifted by 1.7 mag to correct for the larger brightness of young clusters with a mean age of $\sim 10$\,Myr (see Section~\ref{sec:cluster_selection} for caveats of this approach). 

The derived physical properties of the stellar clusters reveal a diverse population, highlighting distinct formation epochs and evolutionary histories within the galaxy. The young star-forming clusters exhibit young ages ($t_M \sim 8$--$11$\,Myr) and stellar masses typically in the range of $\log(M_{\star}/M_{\odot}) \sim 4.5$--$5.5$. These clusters are significantly more metal-rich than the field stars, with median metallicities of $[M/H] \sim -0.3$\,dex, and show moderate dust attenuation ($A_V \sim 0.3$\,mag). These properties are exemplified by Star Cluster 19, displayed in Figure~\ref{fig:starcluster_fits}. Notably, the stellar population models used assume solar-scaled abundances, i.e. [M/H] = [Fe/H].

In contrast, the older GCs display ages consistent with the bulk stellar population of the galaxy, approximately $8.4$--$9.4$\,Gyr, and stellar masses around $\log(M_{\star}/M_{\odot}) \sim5.6$--$5.9$ (exemplified by GC 10 in Figure~\ref{fig:starcluster_fits}). These clusters are significantly more metal-poor, with $[M/H]$ values ranging from $-1.5$ to $-1.3$\,dex, and likely formed during the early assembly phase of GAMA~526784, likely representing remnants of the galaxy’s pre-interaction state. The properties of all clusters obtained with \texttt{BAGPIPES} are summarised in Table~\ref{tab:stellarpops}. 

We were unable to confirm the membership of the old cluster candidates with GAMA~526784 (1, 3, 4, 5, 10, 21, 28, and 29) due to their low S/N spectra. For some of these clusters, only stellar masses could be reliably recovered from the fit, as these rely on the shape of the continuum, which is well traced by the broadband photometry. While the majority of the younger clusters were confirmed to be associated with the galaxy via their emission lines, some (11, 13, 15, 16, 17, 22 and 26) are too faint to reliably recover their stellar population parameters, and we therefore report only their stellar masses and ages, which in the case of younger clusters with F275W detections can be well measured. Notably, some of the clusters are not visible in Figure~\ref{fig:starcluster_fits} because of their faintness. We refer the reader to Figure~\ref{fig:GC_selection}, where the star cluster distribution is shown on top of a deeper, higher-resolution HST $F606W+F814W$ image.

\begin{figure}
    \centering
    \includegraphics[width=\columnwidth]{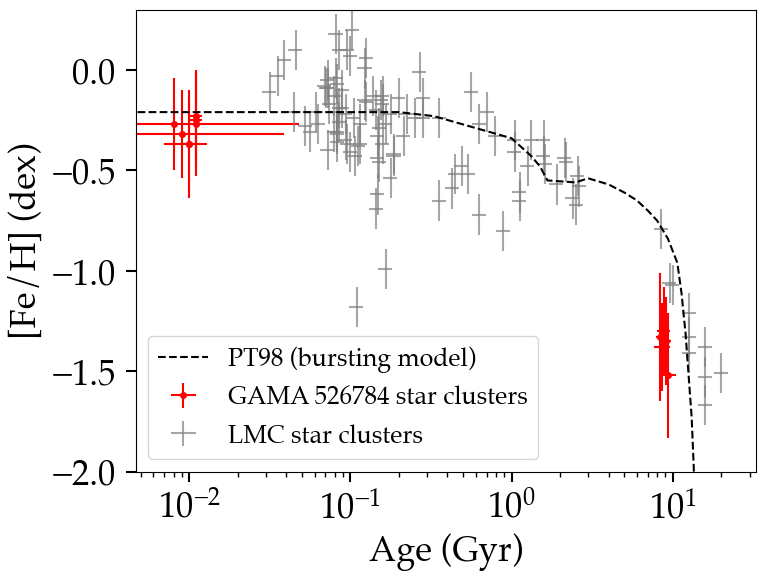}
    \caption{Age-metallicity relation (AMR) for star clusters in GAMA~526784 (red) compared to the star cluster population in the Large Magellanic Cloud (grey scatter points), along with the bursting model of \cite{Pagel_Tautvaisiene_98} to explain the LMC cluster population. The figure illustrates that GAMA~526784 had an evolution similar to that of the LMC, with a prolonged period of quiescence followed by a recent burst of star cluster formation.}
    \label{fig:AMR}
\end{figure}

The star clusters in GAMA~526784 exhibit a clear bimodal distribution, with one population being old and metal-poor and another young and metal-rich. This pattern is similar to the populations of star clusters observed in the Large Magellanic Cloud \citep[LMC,][and references therein]{Narloch_22}.  For a more clear comparison, Figure~\ref{fig:AMR} presents the age-metallicity relation (AMR) of the stellar clusters in GAMA~526784 compared to the star cluster population in the LMC \citep{Narloch_22} (fitted by the bursting model of \citealt{Pagel_Tautvaisiene_98}). Although GAMA~526784 and the LMC are in different stellar mass ranges, both show a complex star formation history of the stellar cluster system, with a significant age gap of $\sim 10$ Gyr between the two main cluster populations. We see a population of clusters in GAMA~526784 which formed 10 Gyr ago, with metallicities of approximately $-1.5$ dex. This was followed by a long phase of stagnation (similar to that observed in the LMC, \citealt{Narloch_22}), during which no significant star cluster formation occurred. This gap lasted from approximately 8 Gyr to 100 Myr ago. A subsequent burst of young cluster formation ignited $\sim$10 Myr ago, producing clusters with metal abundances of approximately $-0.35$ dex.

Estimating the total number of stellar clusters in this galaxy remains challenging, as previously discussed, because the young clusters follow a power-law distribution and the lower-mass ones are likely to dissolve over time. This makes it difficult to predict with certainty whether the galaxy will ultimately turn into a GC-rich UDG. However, if we focus only on seven old GCs (after correcting the detected eight for the expected background contamination of 1.3 sources within the $3~R_e$ FoV, see Appendix \ref{sec:number_density}) likely associated with the central stellar body ($\log(M_{\star}/M_\odot) \sim 8.21$), we find a GC-to-stellar mass ratio of $M_{\rm GC}/M_{\star} = 2.5\%$.
When we extend the analysis to the total stellar mass of the galaxy ($\log(M_{\star}/M_\odot) \sim 8.34$), i.e., including both the central and outer components, the GC mass fraction drops to $M_{\rm GC}/M_{\star} = 1.9\%$. While still elevated compared to typical dwarf galaxies, this value falls short of the 2.5\% threshold often used to classify UDGs as GC-rich \citep{Forbes_25}. However, this estimate is based on the currently detected old clusters and does not apply any completeness correction. Assuming that the seven observed old GCs populate the bright half of the GCLF (as it can be seen in the inset panel in Figure~\ref{fig:starcluster_fits}), the total number of old clusters could plausibly double to 14, likely raising the GC mass fraction to over $2.5\%$, thus configuring a GC-rich UDG without even taking into account the young clusters.

Among the young clusters, eight have masses above the canonical average GC mass ($\log(M_{\star}/M_\odot) \sim 5.3$) and would likely remain above the GCLF turnover as they age. Accounting for the symmetry of the GCLF, these add roughly 16 clusters to the total star cluster population. Therefore, when considering the corrected old GC count and the potential contribution from surviving young massive clusters, the galaxy would host at least 30 clusters. Importantly, the 2.5\% threshold is an arbitrary choice made to find a reasonable separation between the populations of puffed-up dwarfs and failed galaxies by \cite{Forbes_25}, but a significant scatter is expected in this GC mass fraction.

Finally, although some young clusters will certainly be lost through disruption or dissolution effects, the shallow gravitational potential and relative isolation of this galaxy suggest that dissolution effects may not be severe. Consequently, we expect a significant fraction of the massive young clusters to survive, further supporting the conclusion that GAMA~526784 is likely to evolve into a GC-rich UDG over the next several Gyr.

\section{Discussion}
\label{sec:discussion}

GAMA~526784 provides valuable insights into the formation and evolution of UDGs, particularly in the context of isolated star-forming UDGs. In terms of colour \citep{Trujillo_17,Roman_Trujillo_17}, morphology \citep{Buzzo_24, Buzzo_25a, Pfeffer_24}, inclination \citep{ManceraPina_19,ManceraPina_19b}, metallicity \citep{Ferre-Mateu_23,Ferre-Mateu_25,Kado-Fong_24b}, and SFR \citep{Whitaker_12}, the galaxy generally aligns with the expected scaling relations of low-mass galaxies. However, its defining characteristic: a substantial population of young star clusters alongside a modest number of old GCs, contrasts sharply with previous findings, which consistently show that isolated star-forming UDGs barely have any star clusters \citep{Leisman_17, Greco_18, Kadowaki_21, Jones_23}. Notably, this finding of UDGs being GC-poor in the field is as much a function of the environment as it is of the mass of the host \citep{Harris_13}. We note, nonetheless, that there are many massive UDGs in the field that also have few or none GCs, we refer the reader to \cite{Marleau_24} and \cite{Buzzo_24} for some examples.

A direct comparison with the recently studied isolated UDG SMDG0038365-064207 (the ``Disco Ball'', \citealt{Khim_25}) further highlights the distinct nature of GAMA~526784. While both systems host a large population of young star clusters, their spatial distribution and morphology of the host galaxy differ significantly. In the Disco Ball, clusters are uniformly distributed throughout an undisturbed, rotationally-supported stellar disc, with weak, patchy ongoing star formation and no evidence of recent external perturbations. In contrast, GAMA~526784 exhibits a string-like arrangement of star clusters extending over 5 kpc, largely detached from the main stellar body, alongside misaligned gas and stellar kinematics. These features strongly suggest that the cluster formation in GAMA~526784 was triggered by a recent interaction, unlike the likely secular evolution pathway inferred for the Disco Ball.

GAMA~526784’s morphology and kinematics further suggest a complex history. Its two-component structure, with a compact, quiescent inner region and an extended, star-forming outer region, resembles UDGs studied by \citet{Trujillo_17} and \citet{Roman_Trujillo_17}, which often exhibit irregular, clumpy structures indicative of external triggers. The misalignment between the inner and outer components in the ionised gas velocity maps ($\sim 20^{\circ}$) and the elevated velocity dispersion of the ionised gas ($\sigma_{H\alpha} \sim 30$ km s$^{-1}$) resemble features observed in UDGs that have undergone past interactions \citep{Spekkens_18, Leisman_17, Papastergis_17}. These characteristics strongly suggest that GAMA~526784 experienced a perturbation capable of triggering both widespread star formation and the formation of young star clusters.

To explain the presence of these star clusters, we must consider various formation scenarios. One possibility is that GAMA~526784 formed within a high-spin dark matter halo or through a sequence of supernovae feedback, as proposed by \citet{Amorisco_16} and \citet{diCintio_17}. However, the string-like arrangement of clusters and the misaligned gas kinematics suggest that internal processes alone may not fully account for the observed properties.

Alternatively, external interactions may be responsible for the observed properties of GAMA~526784. The high velocity dispersion of the gas ($\sigma_{H\alpha} \sim 30$ km\,s$^{-1}$) may be suggestive of a recent interaction, possibly a flyby or minor merger, which could have compressed the gas and triggered the formation of young star clusters. Similar processes have been proposed for other UDGs with disturbed kinematics \citep{Leisman_17,Spekkens_18}. A particularly intriguing scenario is the bullet dwarf model, which posits that high-speed galaxy collisions can lead to the formation of dark matter-deficient dwarf galaxies with massive star clusters along a trail \citep{Silk_19,vanDokkum_22}. This model predicts that multiple galaxies should form along the collision trail, all with low dark matter content and hosting massive star clusters. Recent simulations by \citet{Lee_24} and observational constraints by \citet{Buzzo_25b} have shown that bullet dwarf-like interactions can produce galaxies with properties similar to GAMA~526784, including misaligned gas kinematics and localised star formation.

While GAMA~526784 shares some similarities with this scenario, such as the presence of young, massive star clusters extending along a string-like region spanning over 5 kpc and evidence of a recent perturbation, it also exhibits key differences. For instance, the lack of a trail of galaxies (as predicted by the bullet dwarf model) and the presence of only a trail of clusters suggest a less extreme interaction. Additionally, the stellar populations of the host galaxy and the clusters differ significantly, which would not be expected if they formed simultaneously during a high-speed collision. The clusters in GAMA~526784 are also less massive than those in other UDGs proposed to result from bullet dwarf collisions, such as DF2 \citep{vanDokkum_18, Danieli_19}, DF4 \citep{vanDokkum_19b, Shen_23}, and FCC 224 \citep{Tang_25, Buzzo_25b}. Furthermore, the stellar velocity dispersion of GAMA~526784 ($\sigma_{\star} = 10$ km\,s$^{-1}$) is too low to conclusively determine its dark matter content, given the nominal resolution of MUSE ($\sim 45$ km\,s$^{-1}$). This uncertainty, combined with the lack of a clear trail of galaxies, makes it difficult to definitively associate GAMA~526784 with the bullet dwarf scenario. Further observations of the galaxy’s neutral and molecular gas content and large-scale environment will be critical to assessing this possibility and clarifying the nature of the possible interaction that shaped its evolution.

Another key formation pathway for GC-rich UDGs is the ``failed galaxy'' scenario, in which an early burst of intense star formation produces numerous GCs before subsequent star formation is suppressed \citep{Danieli_22}. The presence of old GCs and the low stellar metallicity in GAMA~526784 are consistent with this idea. However, the recent resurgence of star formation and the formation of new star clusters present a significant challenge to this model. Classical failed galaxy scenarios predict uniformly old and metal-poor stellar populations with flat metallicity gradients \citep{Ferre-Mateu_25}, as well as a close match between the metallicities of the galaxy and its GCs \citep{Danieli_22, Buzzo_22b}. In contrast, GAMA~526784 exhibits both young and old populations, suggesting a more complex evolutionary pathway than traditional failed galaxy models allow.

Given these possibilities, the most viable explanation involve a high-speed interaction or a sufficiently strong perturbation capable of inducing star cluster formation while leaving the galaxy completely isolated within a few million years. This presents a challenge: either GAMA~526784 experienced an unusually rapid encounter or the galaxy that triggered the star formation was nearly disrupted in the process. To pinpoint the mechanism behind this event, further observational efforts are necessary. Mapping the large-scale environment could reveal possible interaction partners or remnants of past encounters, while resolved molecular and neutral gas studies could help trace interactions and tidal connections. Understanding whether GAMA~526784 represents a rare evolutionary pathway or a missing link in UDG formation models will refine our broader understanding of how these galaxies evolve in isolation.

\section{Conclusions}

We have presented a comprehensive analysis of the UDG GAMA~526784, combining imaging from HST, HSC, and spectroscopy from MUSE. Our study reveals a complex galaxy with a two-component structure: a compact, quiescent inner region and a diffuse, star-forming outer component. The inner component, with an effective radius of $R_{\rm e} \sim 0.9$\,kpc, is dominated by old ($t_M \sim 9.9$\,Gyr), metal-poor ($[M/H] \sim -1.0$) stellar populations, while the outer component, with $R_{\rm e} \sim 2.9$\,kpc, hosts younger ($t_M \sim 0.9$\,Gyr) but slightly more metal-poor ($[M/H] \sim -1.2$) stars. 

The ionised gas analysis reveals low [SII] line ratios (median of 1.1), characteristic of elevated electron densities ($n_e > 400$\,cm$^{-3}$), which may be evidence of ongoing shocks. The gas-phase metallicity is fairly smooth and low (median of 8.3 dex) across the galaxy, with no significant gradients. The total SFR of $0.01 \, M_{\odot} \, \text{yr}^{-1}$ is low but highly localised around the identified stellar clusters. 

Kinematic analysis using \texttt{kinemetry} reveals a clear misalignment between the inner and outer regions in the ionised gas maps of $\sim 20^\circ$. The maps reveal that the stars reach a maximum rotation velocity of 20 km s$^{-1}$, while the gas reaches 40 km s$^{-1}$. The stellar body exhibits a low velocity dispersion ($\sigma_{\star} \sim 10$\,km\,s$^{-1}$). In contrast, the gas component has a much higher velocity dispersion of 30 km  s$^{-1}$, reaching $\sim 50$\,km\,s$^{-1}$ in some regions. 

The star cluster population of GAMA~526784 is particularly intriguing. We identified 29 star clusters, including eight old, metal-poor GC candidates with ages of $\sim 9$\,Gyr and masses of $\log(M_{\star}/M_{\odot}) \sim 5.5$, and 21 young, star-forming clusters with ages of $8$--$11$\,Myr and masses of $\log(M_{\star}/M_{\odot}) \sim 5.0$. The high GC number and high GC mass fraction in the galaxy suggest that GAMA~526784 could be the progenitor of a GC-rich UDG. This makes the galaxy a valuable case study for understanding the early stages of GC formation in UDGs and the conditions under which they evolve into systems with rich GC populations.

The formation history of GAMA~526784 remains complex, with no single scenario fully explaining all its observed properties. The galaxy's two-component structure, kinematic misalignment, and elevated gas velocity dispersion may point to a recent interaction or external perturbation. Some formation scenarios provide partial explanations for its properties. For instance, the bullet dwarf model could account for the formation of young star clusters along a string-like distribution and the apparent isolation of the galaxy. However, the lack of a clear trail of galaxies and the impossibility of constraining the velocity dispersion of the galaxy and, thus, its DM content, make it difficult to fully probe this formation scenario. Alternatively, an interaction with a gas-rich dwarf galaxy could explain the misaligned gas kinematics and the localised star formation, as such interactions are known to compress gas and trigger bursts of star formation. Nevertheless, the presence of both old and young stellar populations suggests that the galaxy's evolution cannot be fully explained by a single event or process. Instead, GAMA~526784 likely experienced a combination of internal and external mechanisms which collectively shaped its current state.

Future observations of the molecular and neutral gas content of GAMA~526784 will provide critical insights into the galaxy's gas reservoirs and large-scale environment. These data will help clarify if the low SII line ratio indeed indicates shocks, assess the availability of gas for future star formation, and determine whether the galaxy's evolution aligns more closely with a bullet dwarf, a minor merger with a gas-rich dwarf, or another pathway entirely. This study highlights the importance of combining photometric, spectroscopic, and kinematic analyses to unravel the complex formation histories of UDGs and their star cluster systems. GAMA~526784 serves as a valuable benchmark for understanding the conditions under which star clusters form in UDGs and their broader implications for dwarf galaxy evolution.

\begin{acknowledgements}
We thank the referee for the important comments and suggestions provided. We deeply thank Duncan Forbes, Rhea Silvia-Remus, Lucas Valenzuela, and Laura Sales for insightful discussions about this study.
This paper is based on observations collected at the European Southern Observatory under ESO programme 108.21ZY.001.
MH acknowledges financial support from the Deutsche Forschungsgemeinschaft (DFG, German Research Foundation) under Germany’s Excellence Strategy – EXC-2094 – 390783311.
AZ acknowledges support from the European Union – NextGeneration EU within PRIN 2022 project n.20229YBSAN - Globular clusters in cosmological simulations and in lensed fields: from their birth to the present epoch and from the INAF Minigrant ‘Clumps at cosmological distance: revealing their formation, nature, and evolution (Ob. Fu. 1.05.23.04.01).
KF acknowledges funding from the European Union’s Horizon 2020 research and innovation programme under the Marie Sk\l{}odowska-Curie grant agreement No 101103830.
\end{acknowledgements}

\bibliographystyle{aa} % style aa.bst
\bibliography{bibli.bib}
\begin{appendix}

\section{Number Density Profile}
\label{sec:number_density}

\begin{figure}
    \centering
    \includegraphics[width=0.9\columnwidth]{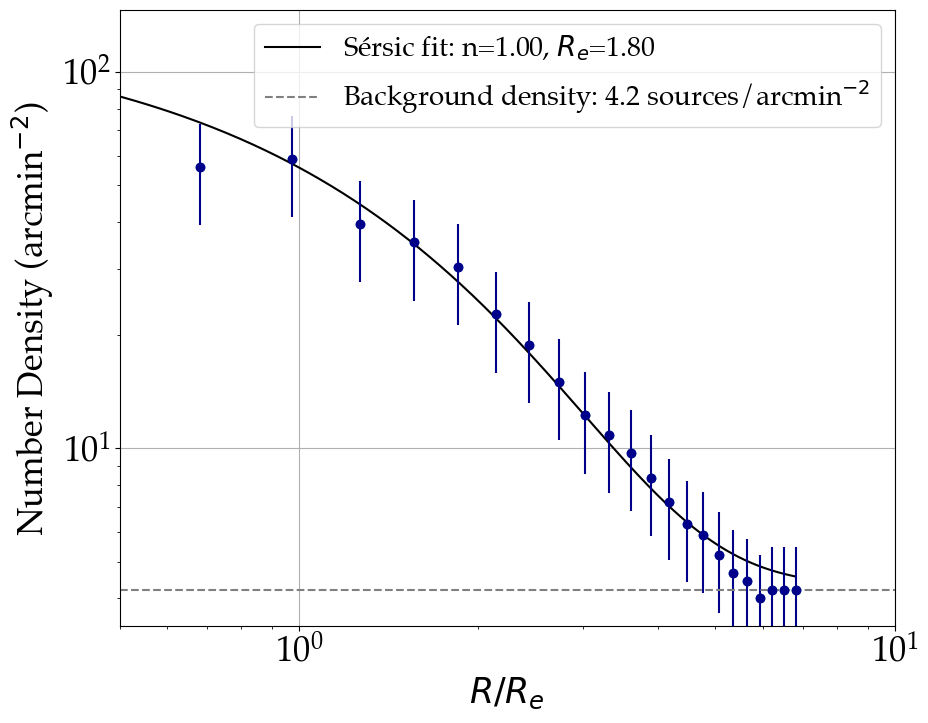}
    \caption{Star cluster surface density profile around GAMA~526784 from galaxy centre in terms of  $R_{\rm e}$. The blue scatter points and error bars are the measurements. The black curve is the S\'ersic model fit to the total star cluster distribution. The dash-dotted grey line is the number density contribution of the background measured outside of 5~$R_{\rm e}$. This number density profile is calculated before selecting the clusters within $3~R_e$ of GAMA~526784.}
    \label{fig:numberdensity}
\end{figure}

We characterised the star cluster system of GAMA~526784 by measuring the radial number density profile of the clusters. The surface density was calculated in equally spaced elliptical bins, from 0.2 to 7 $R_e$, with uncertainties estimated assuming Poisson statistics for the inner bins. A S\'ersic function was fitted to the profile, yielding a S\'ersic index of $n = 1.00 \pm 0.02$ and a star cluster half-number radius in terms of the circularised effective radius is $R_{\rm cluster}/R_{\rm e,circ} = 1.80 \pm 0.32$. The profile is shown in Figure~\ref{fig:numberdensity} The background density was determined both from the fit and from regions beyond 5$R_{\rm e}$, with consistent results of $\rho_{\rm N(bg)} = 4.2 \pm 0.3$ clusters/arcmin$^2$. 
We estimate using this background contamination that within the three effective radii of GAMA~526784 where the star cluster selection was performed, 3.8 interlopers are expected. Out of these, 1.3 are expected to have colours similar to those of old GCs.

\section{Star Formation History}
\label{sec:SFH}
\begin{figure}
\centering
\includegraphics[width=\columnwidth]{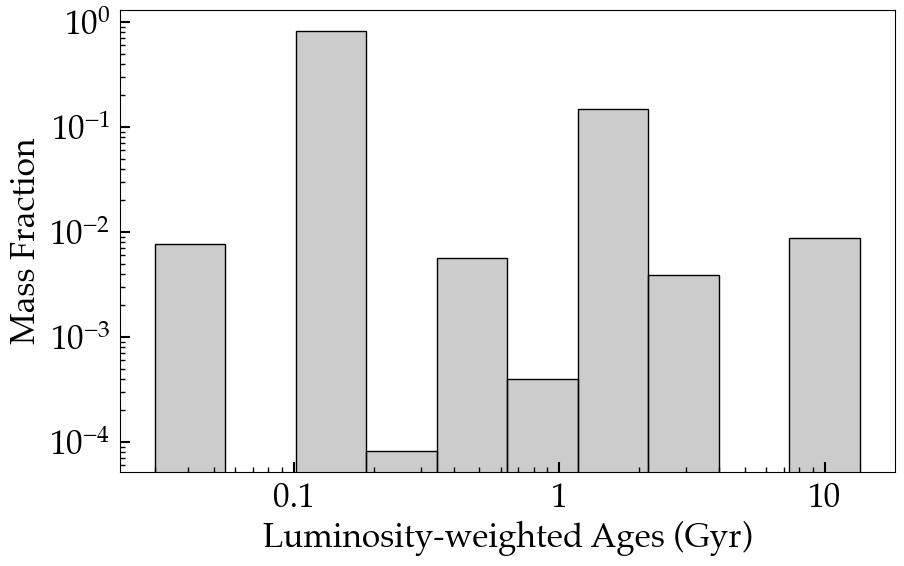}
\caption{Reconstructed star formation history of GAMA~526784, highlighting its extended star formation activity. While the formation of the main stellar body occurred over a prolonged period starting nearly 10 Gyr ago, the SFH is dominated by a more recent and intense burst $\sim$100 Myr ago, which likely led to the formation of massive young star clusters.}
\label{fig:SFH}
\end{figure}

In Figure~\ref{fig:SFH}, we present the reconstructed SFH of GAMA~526784, derived from the stellar population weights measured using \texttt{pPXF}. The SFH reveals a prolonged period of star formation that began nearly 10 Gyr ago, leading to the formation of the galaxy’s main stellar body. Additionally, a more recent, intense burst of star formation occurred approximately 100 Myr ago. Notably, the y-axis of this figure is in logarithmic scale, emphasizing that this recent burst represents the most significant star formation event in GAMA~526784’s history. 

\section{SED fitting and GALFITM results}
\label{sec:appendix_galfitm}

\begin{figure*}
    \centering
    \includegraphics[width=\textwidth]{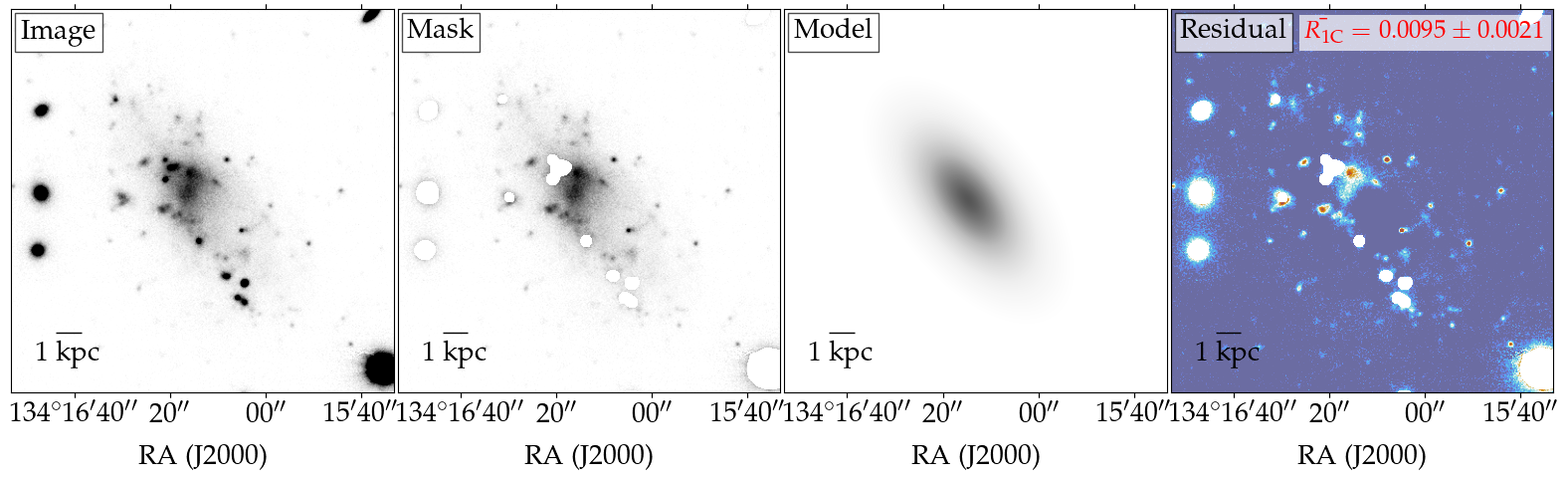}
    \caption{Single S\'ersic \texttt{GALFITM} model of GAMA~526784. \textit{Columns from left to right:} HSC G-band image, segmentation mask from \texttt{SExtractor}, fitted galaxy model with a single component (single-S\'ersic), and residual (image$-$model). The median residuals after removing the contribution from masked sources is $0.0095 \pm 0.0021$, higher than the residuals computed for the two-component model. Clear residuals are observed when only one component is fitted.}
    \label{fig:galfitm_1comp}
\end{figure*}

\begin{figure*}
    \centering
    \begin{subfigure}[t]{0.48\textwidth}
        \centering
        \textbf{Inner Component} % The title above the subfigure
        \par\medskip % Space between the title and the subfigure
        \includegraphics[width=0.98\textwidth]{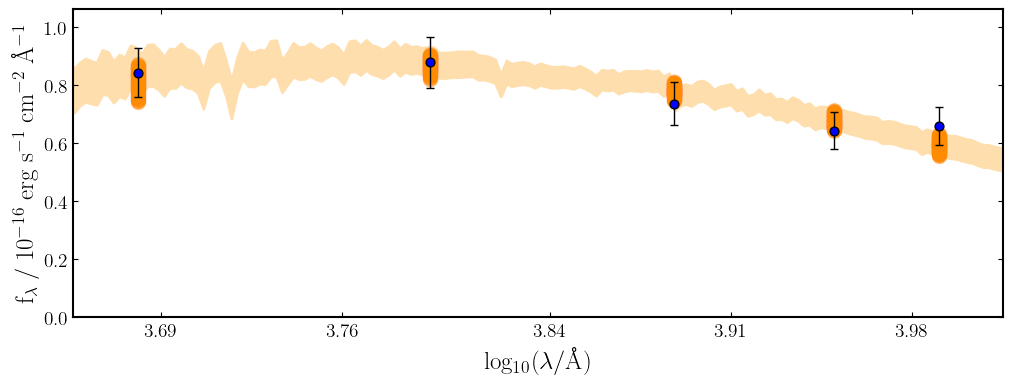}
        \includegraphics[width=0.98\textwidth]{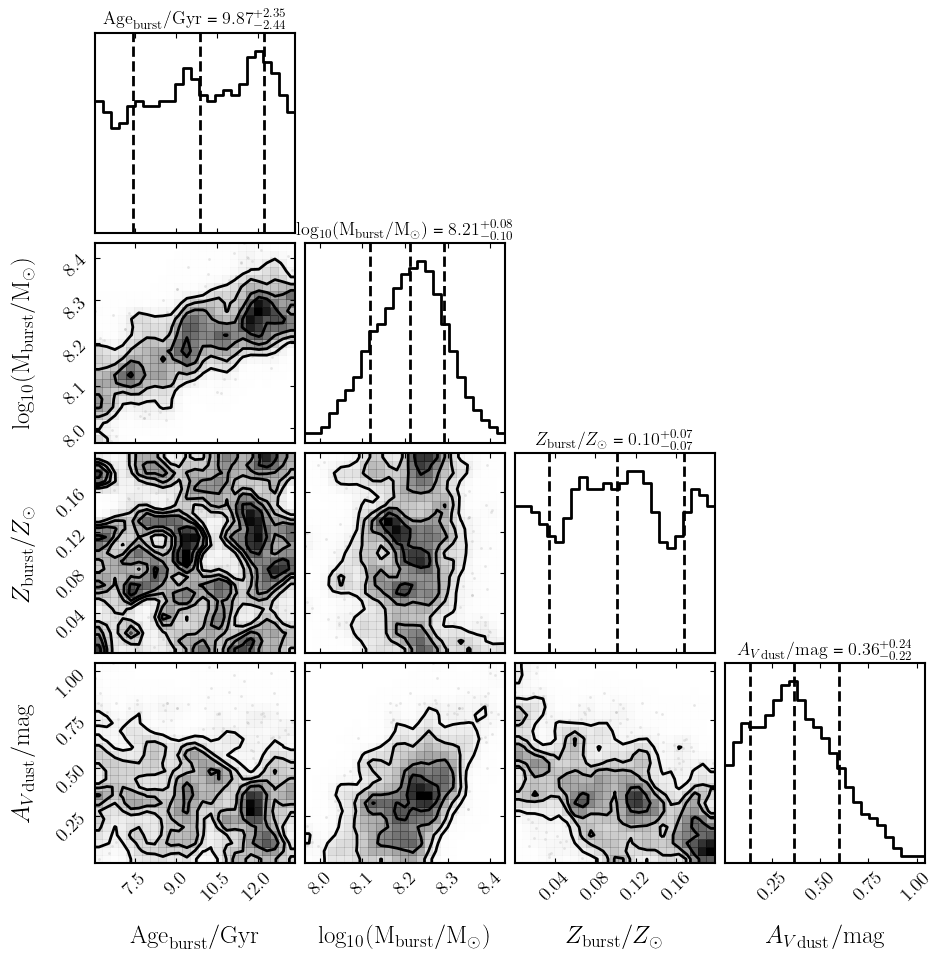}
    \end{subfigure}
    \hfill
    \begin{subfigure}[t]{0.48\textwidth}
        \centering
        \textbf{Outer Component} % The title above the subfigure
        \par\medskip % Space between the title and the subfigure
        \includegraphics[width=0.98\textwidth]{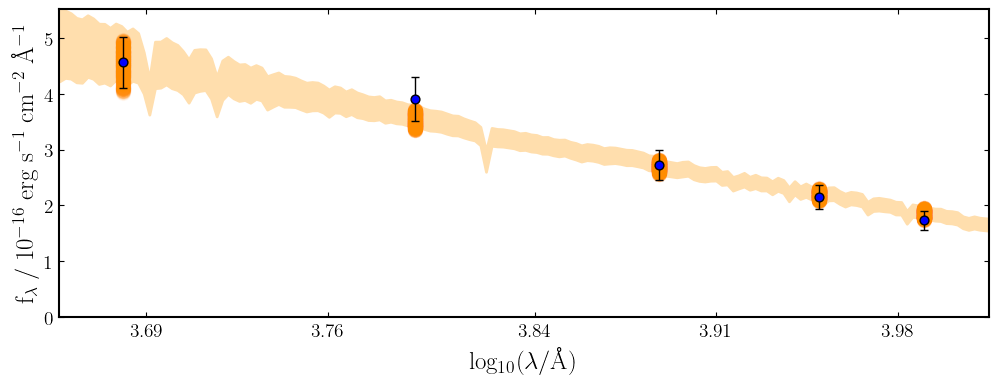}
        \includegraphics[width=0.98\textwidth]{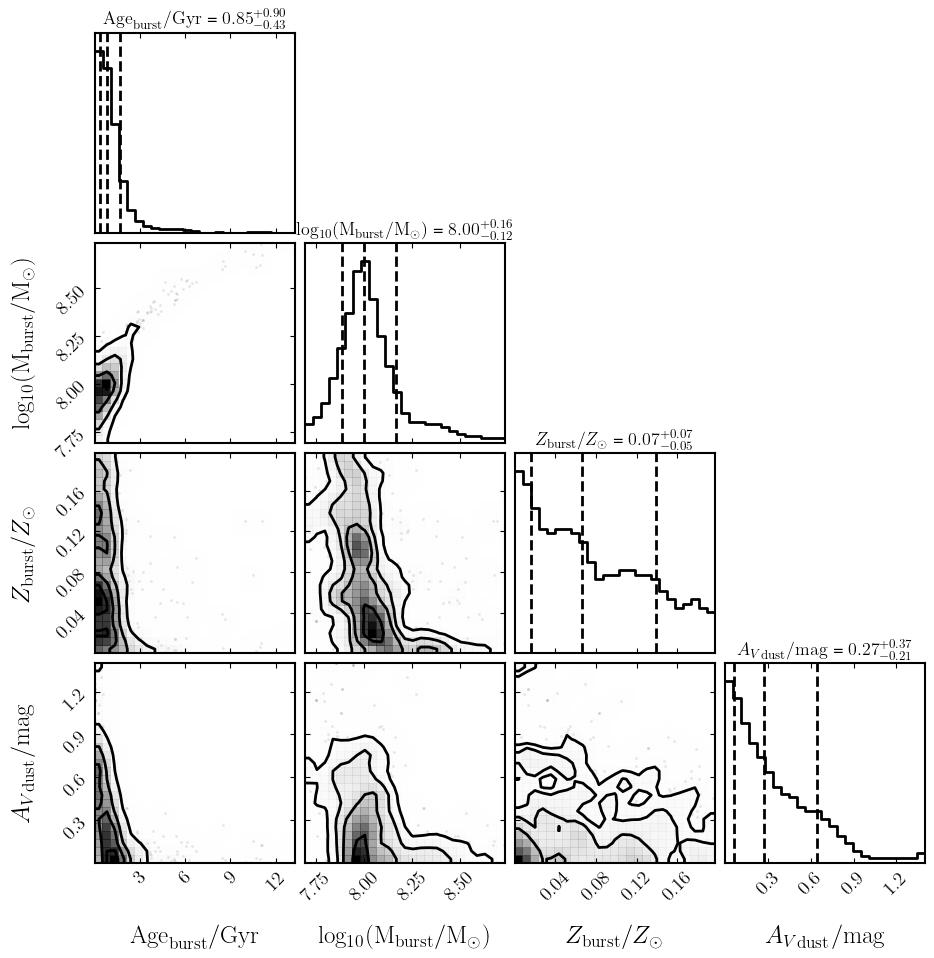}
    \end{subfigure}
    \caption{SED fitting results for the inner (left) and outer (right) components of GAMA~526784. The top panels show the bestfit SED, while the bottom panels show the cornerplot results of the fitted parameters.}
    \label{fig:sed_fitting}
\end{figure*}

In this appendix, we show the result of the \texttt{GALFITM} model for GAMA~526784 using a single S\'ersic component to explain it. This model provides a worse representation of the galaxy, as it can be seen in the residual map which has a higher median value than the model with two components (see Figure~\ref{fig:galfitm}) and has a clear overdensity (i.e. positive residuals) not fit in the centre. The one-component model is shown in Figure~\ref{fig:galfitm_1comp}. The morphological parameters recovered from \texttt{GALFITM} for the two- and one-component models, as well as the stellar population parameters recovered using \texttt{BAGPIPES} for each of them are shown in Table~\ref{tab:data}. For completeness, Figure~\ref{fig:sed_fitting} presents the best-fit SED of GAMA~526784 obtained with \texttt{BAGPIPES} for its two individual components.

\begin{table}

    \centering
    \caption{Results from \texttt{GALFITM} and \texttt{BAGPIPES}}
    \scalebox{0.9}{
    \begin{tabular}{l|ccc} 
        Property & Component 1 & Component 2 & Single S\'ersic \\[0.01cm] \hline \hline
        &  \\[-0.2cm]
        RAJ2000 (deg) & $134.2866$ & $134.2860$ & $134.2859$\\[0.1cm]
        DEJ2000 (deg) & $-0.9755$ & $-0.9765$ & $-0.9762861$\\[0.1cm]
        $g$ (mag) & $19.27 \pm 0.26$ & $17.54 \pm 0.25$ & $17.4 \pm 0.2$\\[0.1cm]
        $g-r$ (mag) & $0.54 \pm 0.15$ & $0.40 \pm 0.23$ & $0.43 \pm 0.15$\\[0.1cm]
        $g-i$ (mag) & $0.79 \pm 0.18$ & $0.47 \pm 0.23$ & $0.53 \pm 0.18$ \\[0.1cm]
        $g-z$ (mag) & $0.95 \pm 0.25$ & $0.52 \pm 0.24$ & $0.58 \pm 0.25$ \\[0.1cm]
        $g-Y$ (mag) & $1.18 \pm 0.23$ & $0.50 \pm 0.35$ & $0.67 \pm 0.23$ \\[0.1cm]
        $R_{\rm e}$ (arcsec) & $5.0 \pm 0.6$ & $15.4 \pm 0.4$ & $13.2 \pm 0.4$\\[0.1cm]
        $R_{\rm e}$ (kpc) & $0.9 \pm 0.1$ & $2.9 \pm 0.1$ & $2.5 \pm 0.1$\\[0.1cm]
        $n$ & $0.57 \pm 0.15$ & $0.46 \pm 0.12$ & $0.67 \pm 0.12$ \\[0.1cm]
        $b/a$ & $0.72 \pm 0.06$ & $0.50 \pm 0.02$ & $0.52 \pm 0.02$ \\[0.1cm]
        $PA$ & $14.2 \pm 1.6$ & $37.6 \pm 0.4$ & $37.1 \pm 0.4$ \\[0.1cm]
        $\mu_{g,0}$ (mag. asec$^{-2}$) & $24.0 \pm 0.3$ & $24.4 \pm 0.2$ & $23.7 \pm 0.3$ \\[0.1cm]
        $\mu_{g,e}$ (mag. asec$^{-2}$) & $24.4 \pm 0.2$ & $24.7 \pm 0.2$ & $24.3 \pm 0.2$ \\[0.1cm]
        log($M_{\star}/M_{\odot}$) & $8.21^{+0.08}_{-0.11}$ & $8.00^{+0.16}_{-0.12}$ & $8.34^{+0.12}_{-0.23}$ \\[0.1cm]
        $[M/H]$ (dex) & $-1.00^{+0.23}_{-0.52}$ & $-1.15^{+0.30}_{-0.55}$ & $-1.12^{+0.47}_{-0.33}$ \\[0.1cm]
        Age (Gyr) & $9.9^{+2.4}_{-2.4}$ & $0.9^{+0.9}_{-0.4}$ & $7.3^{+2.9}_{-4.3}$ \\[0.1cm]
        $A_V$ (mag) & $0.36^{+0.324}_{-0.22}$ & $0.27^{+0.37}_{-0.21}$ & $0.26^{+0.27}_{-0.18}$ \\[0.1cm] \hline
        &  \\[-0.2cm]
        Dist (Mpc) & \multicolumn{3}{c}{$39.0$} \\[0.1cm]
        V (km s$^{-1}$) & \multicolumn{3}{c}{$2758.0 \pm 4.3$} \\[0.1cm] \hline
    \end{tabular}
    \label{tab:data}}
\end{table}

\section{Properties of star clusters}

In this appendix, we present all of the stellar populations results obtained for all star clusters with \texttt{BAGPIPES}.

\begin{table*}
\scalebox{0.95}{
\begin{threeparttable}
\caption{\texttt{BAGPIPES} stellar population properties of the star clusters around GAMA~526784.}
\begin{tabular}{lcccccccccccccccccccc} \hline
\multirow{2}{*}{ID} & RA & Dec & F814W & F606W-F814W & V & \multirow{2}{*}{log($M_{\star}/M_{\odot}$)} & Age  & [M/H] & $A_v$ \\ 
  & [deg] & [deg] & [mag] & [mag] & [km s$^{-1}$]  & & [Gyr] & [dex] & [mag] \\ \hline
1 & $134.2790$	& $-0.9820$ & $22.19 \pm 0.18$ & $0.6519$ & -- & $5.64^{+0.04}_{-0.05}$ & $8.80^{+0.80}_{-1.46}$ & $-1.30^{+0.22}_{-0.24}$ & $0.03^{+0.05}_{-0.02}$ \\
2 & $134.2807$ & $-0.9783$ & $22.96 \pm 0.11$ & $-0.2643$ & $2796.5 \pm 3.5$ & $5.19^{+0.04}_{-0.05}$ & $0.011^{+0.001}_{-0.001}$ & $-0.25^{+0.25}_{-0.24}$ & $0.29^{+0.08}_{-0.08}$ \\
3 & $134.2835$ & $-0.9742$ & $21.77 \pm 0.18$ &	$0.6273$ & -- & $5.90^{+0.04}_{-0.02}$ & $8.60^{+1.00}_{-1.42}$ & $-1.38^{+0.22}_{-0.34}$ & $0.05^{+0.06}_{-0.04}$ \\
4 & $134.2846$ & $-0.9702$ & $23.66 \pm 0.23$ &	$0.5583$ & -- & $5.93^{+0.05}_{-0.05}$ & -- & -- & -- \\
5 & $134.2820$ & $-0.9818$ & $23.24 \pm 0.20$ &	$0.3079$ & -- & $5.89^{+0.04}_{-0.05}$ & $9.03^{+0.72}_{-1.34}$ & $-1.35^{+0.22}_{-0.24}$ & $0.14^{+0.22}_{-0.23}$ \\
6 & $134.2847$ & $-0.9743$ & $21.52 \pm 0.14$ & $0.4412$ & $2764.6 \pm 4.7$ & $5.75^{+0.02}_{-0.14}$ & $0.009^{+0.002}_{-0.002}$ & $-0.42^{+0.05}_{-0.17}$ & $0.48^{+0.02}_{-0.03}$ \\
7 & $134.2831$ & $-0.9818$ & $23.04 \pm 0.15$ &	$0.1345$ & $2815.1 \pm 3.3$ & $4.88^{+0.12}_{-0.16}$ & $0.007^{+0.001}_{-0.002}$ & $-0.43^{+0.22}_{-0.14}$ & $0.50^{+0.01}_{-0.01}$ \\
8 & $134.2841$ & $-0.9810$ & $22.11 \pm 0.12$ & $-0.0112$ & $2812.1 \pm 5.4$& $5.33^{+0.03}_{-0.03}$ & $0.008^{+0.002}_{-0.003}$ & $-0.33^{+0.12}_{-0.05}$ & $0.46^{+0.03}_{-0.05}$\\
9 & $134.2837$ & $-0.9812$ & $23.13 \pm 0.09$ &$0.3462$ & $2797.8 \pm 3.7$ &  $5.06^{+0.02}_{-0.02}$ & $0.010^{+0.001}_{-0.003}$ & $-0.34^{+0.05}_{-0.07}$ & $0.49^{+0.01}_{-0.02}$\\
10 & $134.2851$ & $-0.9755$ & $22.96 \pm 0.24$ &	$0.5440$ & -- & $5.77^{+0.07}_{-0.09}$ & $8.28^{+0.44}_{-0.21}$ & $-1.33^{+0.32}_{-0.12}$ & $0.02^{+0.03}_{-0.01}$ \\
11 & $134.2835$ & $-0.9819$ & $23.35 \pm 0.12$ &	$-0.0879$ & $2814.7 \pm 5.1$ & $4.50^{+0.14}_{-0.06}$ & $0.013^{+0.003}_{-0.004}$ & -- & -- \\
12 & $134.2841$ & $-0.9808$ & $23.69 \pm 0.15$ &	$0.4848$ & $2809.7 \pm 4.3$ & $5.50^{+0.05}_{-0.07}$ & $0.011^{+0.001}_{-0.001}$ & $-0.27^{+0.26}_{-0.26}$ & $0.42^{+0.06}_{-0.10}$ \\
13 & $134.2859$ & $-0.9734$ & $23.04 \pm 0.19$ &	$0.4451$ & -- & $5.49^{+0.12}_{-0.16}$ & $0.010^{+0.005}_{-0.003}$ & -- & -- \\
14 & $134.2847$ & $-0.9790$ & $23.18 \pm 0.11$ &	$0.1789$ & $2796.1 \pm 5.2$ & $4.60^{+0.06}_{-0.05}$ & $0.010^{+0.004}_{-0.004}$ & $-0.37^{+0.12}_{-0.10}$ & $0.37^{+0.09}_{-0.10}$ \\
15 & $134.2848$ & $-0.9797$ & $23.59 \pm 0.13$ &	$0.1911$ &  $2795.7 \pm 5.2$ & $4.93^{+0.13}_{-0.14}$ & $0.010^{+0.002}_{-0.002}$ & -- & --\\
16 & $134.2865$ & $-0.9734$ & $23.52 \pm 0.21$ &	$0.4798$ & -- & $5.73^{+0.03}_{-0.13}$ & $0.008^{+0.004}_{-0.003}$ & -- & -- \\ 
17 & $134.2867$ & $-0.9735$ & $23.06 \pm 0.13$ &	$0.5600$ & $2729.6 \pm 6.7$ & $4.17^{+0.12}_{-0.05}$ & -- & -- & -- \\
18 & $134.2864$ & $-0.9773$ & $22.54 \pm 0.13$ &$0.2589$ & $2786 \pm 7$ & $5.69^{+0.09}_{-0.09}$ & $0.007^{+0.002}_{-0.003}$ & $-0.23^{+0.03}_{-0.17}$ & $0.39^{+0.11}_{-0.07}$ \\
19 & $134.2861$ & $-0.9782$ & $21.66 \pm 0.12$ &	$0.2948$ & $2786.4 \pm 2.4$ & $5.52^{+0.05}_{-0.06}$ & $0.010^{+0.001}_{-0.002}$ & $-0.37^{+0.23}_{-0.27}$ & $0.45^{+0.04}_{-0.07}$ \\ 
20 & $134.2860$ & $-0.9783$ & $23.74 \pm 0.19$ &	$0.4273$ & $2793.1 \pm 4.7$ & $5.47^{+0.02}_{-0.02}$ & $0.007^{+0.005}_{-0.003}$ & $-0.27^{+0.03}_{-0.17}$ & $0.29^{+0.04}_{-0.02}$ \\
21 & $134.2861$ & $-0.9782$ & $21.82 \pm 0.26$ &	$0.2755$ & -- & $5.69^{+0.01}_{-0.01}$ & $8.45^{+0.30}_{-0.30}$ & $-1.43^{+0.06}_{-0.12}$ & $0.03^{+0.05}_{-0.02}$  \\
22 & $134.2875$ & $-0.9728$ & $23.05 \pm 0.14$ &$0.4510$ & -- & $4.27^{+0.32}_{-0.04}$ & $0.007^{+0.03}_{-0.01}$ & -- & -- \\
23 & $134.2877$ & $-0.9743$ & $22.10 \pm 0.15$ &	$-0.1555$ & $2741.2 \pm 4.4$ & $5.05^{+0.04}_{-0.03}$ & $0.008^{+0.04}_{-0.02}$ & $-0.32^{+0.21}_{-0.27}$ & $0.48^{+0.01}_{-0.02}$ \\
24 & $134.2870$ & $-0.9769$ & $23.12 \pm 0.17$ &	$0.5716$ & $2771.3 \pm 5.6$& $4.55^{+0.03}_{-0.03}$ & $0.010^{+0.03}_{-0.04}$ & $-0.27^{+0.23}_{-0.27}$ & $0.48^{+0.01}_{-0.02}$\\ 
25 & $134.2869$ & $-0.9768$ & $23.55 \pm 0.12$ &	$0.6368$ & $2776.4 \pm 7.2$ & $4.58^{+0.05}_{-0.03}$ & $0.010^{+0.02}_{-0.01}$ & $-0.33^{+0.13}_{-0.17}$ & $0.48^{+0.01}_{-0.03}$ \\
26 & $134.2881$ & $-0.9768$ & $23.34 \pm 0.14$ &	$0.4408$ & $2757.5 \pm 4.8$ &$4.91^{+0.03}_{-0.03}$ & $0.011^{+0.001}_{-0.001}$ & -- & -- \\
27 & $134.2901$ & $-0.9712$ & $23.69 \pm 0.08$ &	$-0.1530$ & $2730 .1 \pm 3.2$ & $4.65^{+0.05}_{-0.07}$ & $0.007^{+0.003}_{-0.002}$ & $-0.33^{+0.05}_{-0.12}$ & $0.36^{+0.08}_{-0.07}$  \\
28 & $134.2901$ & $-0.9704$ & $23.64 \pm 0.19$ &	$0.2360$ & -- & $5.50^{+0.15}_{-0.04}$ & -- & -- & -- \\
29 & $134.2896$ & $-0.9775$ & $23.29 \pm 0.23$ &	$0.2840$ & -- & $5.83^{+0.04}_{-0.07}$ & $8.88^{+1.73}_{-0.59}$ & $-1.37^{+0.23}_{-0.27}$ & $0.02^{+0.04}_{-0.02}$ \\ \hline
\end{tabular}
\begin{tablenotes}
      \small
      \item \textbf{Note.} Columns are: (1) ID; (2) Right Ascension; (3) Declination; (4) F814W magnitude; (5) F606W-F814W colour; (5) \texttt{pPXF} radial velocity; (6) \texttt{BAGPIPES} stellar mass; (7)  \texttt{BAGPIPES} Age; (8) \texttt{BAGPIPES} metallicity; (9) \texttt{BAGPIPES} Dust attenuation.
\end{tablenotes}
\label{tab:stellarpops}
\end{threeparttable}}
\end{table*}
\end{appendix}

\end{document}